# Epitaxial engineering of flat silver fluorides cuprate analogs

*Adam Grzelak[1]\*, Haibin Su[2]\*, Xiaoping Yang[3], Dominik Kurzydłowski[1,4], José Lorenzana[5]\*, and Wojciech Grochala[1]\**

[1]*Center of New Technologies, University of Warsaw, 02089 Warsaw, Poland*
[2]*Department of Chemistry, Hong Kong University of Science and Technology, Hong Kong, Hong Kong*
[3]*Anhui Province Key Laboratory of Condensed Matter Physics at Extreme Conditions, High Magnetic Field Laboratory, Chinese Academy of Sciences, Hefei 230031, China*
[4]*Faculty of Mathematics and Natural Sciences, Cardinal Stefan Wyszyński University in Warsaw, 01938 Warsaw, Poland*
[5]*Institute for Complex Systems (ISC), Consiglio Nazionale delle Ricerche, Dipartimento di Fisica, Università di Roma "La Sapienza", 00185 Rome, Italy*
*\*a.grzelak@cent.uw.edu.pl, haibinsu@ust.hk, jose.lorenzana@cnr.it, w.grochala@cent.uw.edu.pl*

**Abstract**

As-grown $AgF_2$ has a remarkably similar electronic structure as insulating cuprates, but it is extremely electronegative, which makes it hard to handle and dope. Furthermore, buckling of layers reduces magnetic interactions and enhances unwanted self-trapping lattice effects. We argue that epitaxial engineering can solve all these problems. By using a high throughput approach and first principle computations, we find a set of candidate substrates which can sustain the chemical aggressiveness of $AgF_2$ and at the same time have good lattice parameter matching for heteroepitaxy, enhancing $AgF_2$ magnetic and transport properties and opening the possibility of field-effect carrier injection to achieve a new generation of high-$T_c$ superconductors. Assuming a magnetic mechanism and extrapolating from cuprates we predict that the superconducting critical temperature of a single layer can reach 195 K.

**Introduction**

In recent years, silver(II) fluorides have attracted increased scientific attention because of their similarities to copper oxides, some of which are known as precursors to the most important family of high-temperature superconductors. In particular, silver(II) fluoride $AgF_2$ has been found to exhibit many of the same traits as one such precursor – $La_2CuO_4$. [1] These traits – layered crystal structure and a charge-transfer type correlated insulating state with strong two-dimensional antiferromagnetic (AFM) coupling – are thought to be crucial for high-$T_c$ superconductivity in cuprates, which makes $AgF_2$ and its derivatives a promising area of further study. Very recently Miller and Botana further highlighted the similarities between $AgF_2$ and oxocuprates in terms of electronic and magnetic structures. [2]

It has been shown that buckling of Ag-F-Ag bonds within $AgF_2$ layers prevents the magnetic coupling constant (*J*) from reaching values as high as those found in oxocuprates and reduces the bandwidth, favoring lattice polaronic effects. [1] Indeed, in a puckered structure the out-of-plane ligand mode couples with the electronic structure to first order in displacements, while in a flat structure it couples to second order, strongly reducing polaronic effects. [3]

There have been some previous experimental attempts at obtaining flat layers of $AgF_2$, as well as several computational studies exploring this possibility. Ternary fluoroargentates(II), such as $KAgF_3$ or Cs$_2$AgF$_4$, contain flat layers in their crystal structures, but antiferro orbital ordering accompanied by antiferrodistortive ordering of Ag-F bonds (*i.e.* that the bonds are alternately long and short) within any given layer precludes the emergence of two-dimensional (2D) AFM coupling [4–6] according to Goodenough-Kanamori-Anderson rules. [7,8]

Initial computational studies suggested that a flat-layered AgF$_2$ polymorph could be stable at high pressure [9] and exhibit much enhanced superexchange interactions as compared to ambient pressure polymorph; [10] however, subsequent X-ray diffraction studies of AgF$_2$ compressed up to 40 GPa and more detailed DFT+U calculations have shown that AgF$_2$ transforms instead into a nanotubular polymorph at *ca.* 15 GPa. [11] Although these nanotubes lack long-range AFM ordering, interactions within constituent [Ag$_2$F$_7$]$^{3-}$ dimers via almost linear Ag-F-Ag bridges are strongly enhanced (3.5-fold) compared to ambient-pressure AgF$_2$, which further justifies the search for flat-layered AgF$_2$ systems exhibiting linear Ag-F-Ag bridges. [12]

Theoretical models of an isolated layer of AgF$_2$ suggest that it preserves puckering typical of its condensed phase form. [13] Our proposition is to explore the possibility to stabilize a flat form of AgF$_2$ via epitaxial engineering. We are motivated by the enormous progress in this field in the last decades. Similar to our aim, a structure different from the bulk form of CuO has been stabilized with this approach. [14] Furthermore, it has being demonstrated that using a double layer transistor geometry it is possible to explore a sizable part of the phase diagram of a unit cell thick cuprate raising $T_c$ from zero to nearly 30 K. [15] Also encouraging is the order of magnitude enhancement of $T_c$ in FeSe on going from the bulk form to a monolayer grown on SrTiO$_3$, [16,17] the overcome of the maximum $T_c$ of the La$_{2-x}$Sr$_x$CuO$_4$ family at the interface between non-superconducting underdoped and overdoped layers [18] as well as metallicity [19] and superconductivity [15,20] appearing at the interface of insulating oxides.

AgF$_2$ is a commercially available compound, utilized for its strong oxidizing and fluorinating properties. [21] Indeed, recent computations of the work function of AgF$_2$ show that a single layer of this compounds could easily take away electrons from the surface of most inorganic solids which it comes into contact with. [13] Thus, the identification of an appropriate substrate that can satisfy all epitaxial and chemical constraints is extremely challenging. In this work, we show that the list of materials that can serve as substrates or spacers is limited but not at all null and consists of certain fluorides of closed-subshell metal cations. We theoretically explore stability, as well as magnetic and electronic properties of a single flat layer of AgF$_2$ deposited on the most promising candidates (binary and ternary metal fluorides). Using a semiempirical approach we estimate the maximum critical temperature attainable upon doping. In addition, we examined AgF$_2$ monolayer placed on an exemplary oxide surface (MgO) to get insight into a possible electron-transfer between reactive AgF$_2$ monolayer and the substrate.

**Computational methods**

All calculations were carried out within density functional theory (DFT) approach as implemented in VASP software, [22–26] using GGA-type Perdew-Burke-Ernzerhof functional adapted for solids (PBEsol). [27] On-site Coulombic interactions of Ag d electrons were accounted for through DFT+U correction as introduced by Liechtenstein *et al.*, [28] with the Hubbard *U* and Hund $J_H$ parameters for Ag set to 5 eV and 1 eV, respectively. [29] Plane-wave cutoff energy of 520 eV was used in all systems, except for those containing lithium (650 eV). A wave vector spacing of 0.03 Å$^{-1}$ was used for all calculations. NUPDOWN tag was used to enforce total spin in the FM state.

For all candidate substrates, we first optimized the geometry of the bulk compound in a 2x2x2 supercell. In the next step, we created a vacuum slab containing several layers of the compound and relaxed atomic coordinates in the direction perpendicular to the surface (by convention referred to as *z*), except for three central layers. We then placed an AgF$_2$ monolayer on the surface and relaxed *z* coordinates of its atoms and the atoms in the top three layers of the substrate.

Magnetic interactions in the AgF$_2$ layer were evaluated with antiferromagnetic coupling constant $J_{2D}$ ("2D" stands for two-dimensional coupling within the monolayer), calculated using collinear configurations and the broken symmetry method as $J_{2D} = E_{AFM} - E_{FM}$, where $E_{AFM}$ and $E_{FM}$ are energies per silver of antiferromagnetic and ferromagnetic solutions, respectively. By this convention, $J_{2D}$ is negative in AFM systems. In all the discussion below by "larger/smaller $J_{2D}$" we mean larger/smaller in magnitude. For most interesting systems, $J_{2D}$ was additionally calculated with HSE06 hybrid functional. [30] VESTA [31] software was used for visualization of structures, spin density maps and electron localization function. Band structures and electronic density of states (eDOS) were plotted using PyProcar [32] and p4vasp [33] software.

**Results and discussion**

Epitaxial growth of a single layer of AgF$_2$ on a fluoride substrate, which enforces a particular length of Ag-F-Ag bridges, represents an experimental possibility for strain stabilization of a single flat layer. Consequently, we have screened the ICSD structural database for candidate structures and selected seven known fluoride systems based on two criteria: a) **cell dimensions** in the range ca. 4.0–4.2 Å, corresponding to Ag-F distance of 2.0–2.1 Å – experimental value in bulk AgF$_2$ is ca. 2.07 Å [34]; b) **chemical composition** – no open d-subshell metal cations to avoid magnetism and propensity towards oxidation. These fluorides adopt either rock-salt or perovskite structure, while SnF$_4$ constitutes a sublattice of a classical tetragonal double perovskite structure. We also used three hypothetical hybrid fluorides: KZn$_{0.5}$Cd$_{0.5}$F$_3$, KZn$_{0.25}$Cd$_{0.75}$F and Na$_{0.25}$Li$_{0.75}$F, which provided more data points filling the gaps within the aforementioned Ag-F bond length range. On top of that, three systems which did not meet some of the criteria above, were selected for comparison: a) KCdF$_3$ and RbCdF$_3$, with an even larger unit cell vector (*ca.* 4.4 Å) and b) MgO (an oxide rather than a fluoride).

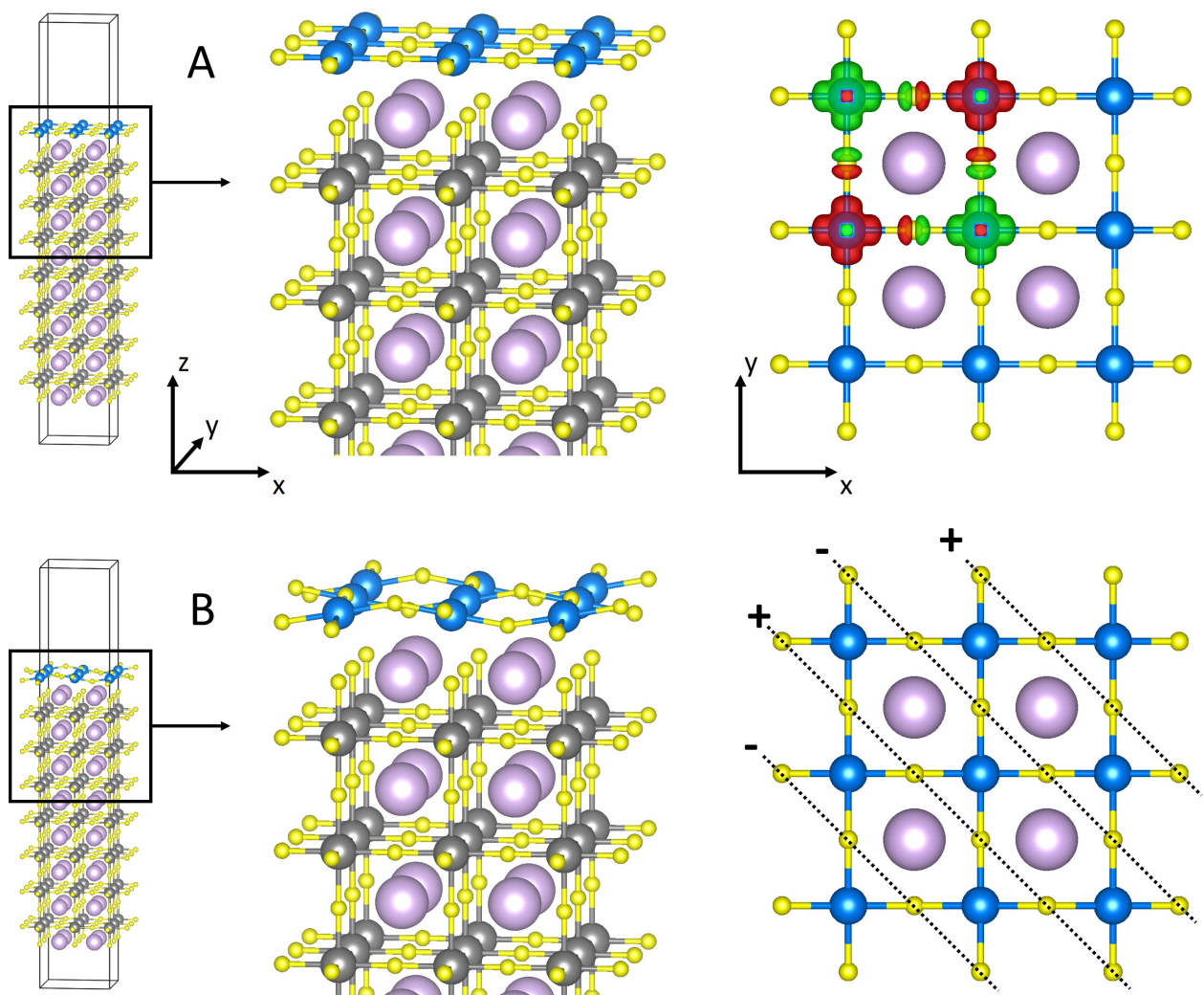

**Figure 1.** AgF$_2$ monolayer on an example substrate (KMgF$_3$). A – tetragonal (flat), B – orthorhombic (corrugated). The "+" and "–" signs indicate *z* displacement of F atoms relative to Ag atoms. Left panel – view of the entire unit cell, central panel– close-up of several top layers, right panel – view along *z* axis. Color code: blue – Ag, yellow – F, purple – K, dark grey – Mg. Ag-F bonds between monolayer and substrate and K-F bonds are not shown for clarity. In the top right panel, we show also the magnetization density in selected sites (red and green).

An example of the optimized AgF$_2$-on-fluoride model is presented in fig. 1. Each silver atom is coordinated by five F atoms: four in the in-layer square unit, similar to constituent units of bulk AgF$_2$, and another one in the apical position, from the fluoride substrate. The five F atoms form a tetragonal pyramid around each Ag atom. The distance to the apical F atom is in all cases larger than to the in-layer F atoms, which can be interpreted as a manifestation of Jahn-Teller effect in a d$^9$ metal cation such as Ag(II). Detailed structural information on AgF$_2$ monolayer on studied fluoride substrates are listed in table 1, and the cif files are given in Supplementary Material.

Table 1. Structural and magnetic data for AgF$_2$ layers on fluoride substrates, calculated with PBEsol+U, listed in the order of increasing Ag-Ag distance for fluoride substrates. FU stands for a formula unit of AgF$_2$.

Footnotes:

*E(g.s.) is the energy of the lowest energy (ground state) configuration, E(flat) refers to the energy of the almost flat (i.e. Ag-F-Ag angle > 175 deg.) tetragonal configuration with all Ag-F bonds equal, obtained by starting from a perfectly flat AgF$_2$ layer. In regions I and III, the lowest energy configurations are orthorhombic: corrugated and quasi-molecular, respectively. In region II, the solutions starting from either corrugated or quasi-molecular structures lead to almost flat structure, equivalent in energy to the tetragonal solution; thus, the flat (tetragonal) solution is the ground state.

†In region I, the average value of Ag-F bond length is given; in region III, the shorter distance (i.e. within AgF$_2$ quasi-molecule) is given.

In region I structures, there are two non-equivalent Ag-F-Ag angles which differ by less than 3 deg. We report the average value.

‡The sign represents the F higher (+) or lower (-) than the Ag in the uppermost layer. +- is the two-up, two-down configurations of the four F's surrounding an Ag as represented in Fig. 1B. ++ or -- represents all F's displaced in the same direction.

§Hypothetical flat polymorph of AgF$_2$.

| | substrate | Ag-Ag distance [Å] | Ag-F bond [Å] | | Ag-F-Ag angle [deg.] | Symmetry‡ | J$_{2D}^{PBEsol+U}$ [meV] | Ag magnetic moment [μ$_B$] | E(g.s.)–E(flat)* [meV/FU] |
|---|---|---|---|---|---|---|---|---|---|
| | | | in-layer† | apical | | | | | |
| I | KMgF$_3$ | 3.985 | 2.025 | 2.434 | 159.3 | +- | -214.9 | 0.503 | -29.6 |
| | AgZnF$_3$ | 3.989 | 2.046 | 2.380 | 154.0 | +- | -174.9 | 0.510 | -73.7 |
| | LiF | 3.999 | 2.041 | 2.544 | 156.1 | +- | -194.6 | 0.506 | -59.5 |
| II | RbMgF$_3$ | 4.055 | 2.028 | 2.414 | 178.4 | ++ | -264.7 | 0.486 | 0.0 |
| | KZnF$_3$ | 4.059 | 2.029 | 2.373 | 178.0 | -- | -234.1 | 0.469 | 0.0 |
| | SnF$_4$ | 4.106 | 2.053 | 2.715 | 179.5 | ++ | -250.6 | 0.477 | 0.0 |
| | CsMgF$_3$ | 4.177 | 2.089 | 2.460 | 175.2 | ++ | -207.0 | 0.486 | 0.0 |
| | Na$_{0.25}$Li$_{0.75}$F | 4.217 | 2.109 | 2.391 | 175.8 | -- | -207.8 | 0.479 | 0.0 |
| | KZn$_{0.5}$Cd$_{0.5}$F$_3$ | 4.226 | 2.113 | 2.299 | 176.9 | -- | -150.1 | 0.439 | 0.0 |
| III | KZn$_{0.25}$Cd$_{0.75}$F$_3$ | 4.315 | 2.054 | 2.219 | 176.9 | -- | -74.2 | 0.457 | -11.5 |
| | KCdF$_3$ | 4.399 | 2.034 | 2.164 | 177.2 | -- | -49.5 | 0.481 | -52.3 |
| | RbCdF$_3$ | 4.422 | 2.029 | 2.176 | 179.0 | -- | -55.0 | 0.493 | -87.1 |
| | MgO | 4.210 | 2.117 | 2.410 | 167.8 | -- | +18.2 | 0.182 | - |
| | Bulk AgF$_2$ (tetragonal)§ | 4.046 | 2.023 | n/a | 180.0 | +- | -288.1 | 0.487 | n/a |
| | Bulk AgF$_2$ (orthorhombic) | 3.736 | 2.070 | 2.569 | 129.0 | n/a | -51.0 | 0.570 | -292.9 |

Since an isolated layer of AgF$_2$ has a tendency to corrugate, [13] we also performed optimization of an initially corrugated structure on all fluoride surfaces listed in table 1. Overall, we can distinguish three domains in the studied range of Ag-Ag distance: I – below *ca.* 4.0 Å (too short), II – 4.0-4.25 Å (optimum), III – above *ca.* 4.25 Å (too long), and the behavior of the AgF$_2$ layer is different in each domain. For substrates with unit cell dimensions in **domain I**, the orthorhombic corrugated solution is more stable. Above a certain Ag-Ag distance (ca. 4.0 Å) (**domain II**), initially corrugated layer flattens out during optimization and results in virtually the same geometry as optimization of flat layer – the Ag-F-Ag bond angle differs by no more than 0.2° in most of these cases. Depending on the substrate, we have found cases in which the energy gain due to corrugation of a flat layer is quite large (tens of meV per formula unit (FU) of AgF$_2$, which corresponds to several kJ/mol) and cases in which there is no tetragonal distortion. Strikingly, we have found no cases in between suggesting that the transition between the flat tetragonal (**domain I**) and corrugated orthorhombic (**domain II**) is quite abrupt.

The ground state of the flat layer in **domain II** is an antiferromagnetic charge-transfer insulator according to the Zaanen-Sawatzky-Allen classification scheme, [35] as discussed in more detail below. The magnetic moment is due to a half-filled d($x^2$-$y^2$) orbital nicely mimicking cuprates as can be seen from the shape of the magnetization density in Fig. 1.

Finally, while AgF$_2$ layer in **domain III** is flat, as desired, yet the tetragonal arrangement (typical for domain II) is now unstable towards symmetry lowering. Examination of the magnetization density reveals that the orbital ordering is different. Indeed, instead of the half-filled d($x^2$-$y^2$) orbital now a d($z'^2$)-like orbital sustains the magnetization, with z' oriented along the planar Ag-F bond and rotating from one site to the next. Clearly the orbital order has switched to antiferro and the resulting geometry is composed of quasi-0D AgF$_2$ dumbbells where each Ag atom has two shorter and two longer F contacts (fig. 2). This implies the presence of both short and long Ag-F bonds in the superexchange pathway. Typically, these solutions turned out to be more stable by between 10 to 90 meV/FU i.e. several kJ/mol (increasing with unit cell dimensions) from the higher-symmetry solutions in the cases studied.

For comparison, analogous calculations in systems in domain II (RbMgF$_3$, SnF$_4$ and Na$_{0.25}$Li$_{0.75}$F) starting from lower-symmetry quasi-0D AgF$_2$ molecules yielded a flat tetragonal AgF$_2$ monolayer with uniform Ag-F bond length, equivalent to the solution listed in table 1, indicating that systems in domain II (with Ag-Ag distance shorter than 4.3 Å) do not undergo decomposition into 'molecules'.

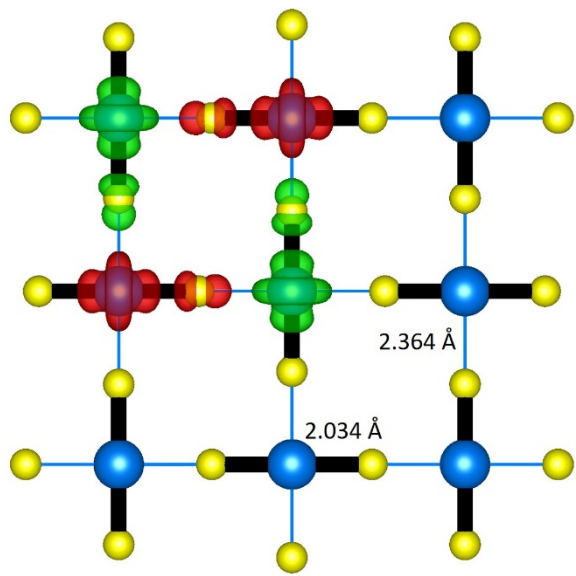

Figure 2. Lowest-energy structure of AgF$_2$ monolayer on sample substrate from domain III – view along *z* axis. Note the varied Ag-F distance (given for KCdF$_3$), indicating AgF$_2$ molecule formation. Red and green show the magnetization density in selected sites.

Since the energy difference between layers with flat and corrugated orientation in bulk AgF$_2$ (table 1) is equal to ca. 293 meV/FU (ca. 28 kJ/mol), one may conclude that epitaxial deposition of AgF$_2$ on proper substrates typical of domain II indeed has the potential of overcoming the substantial energy barrier needed for obtaining stable, flat layers of AgF$_2$. In addition, we investigated the nature of bonding between AgF$_2$ monolayer and fluoride substrate by means of electron localization function (ELF) visualization. [36] The ELF analysis indicates predominantly ionic interactions between the AgF$_2$ layer and the substrate; the detailed results can be found in ESI.

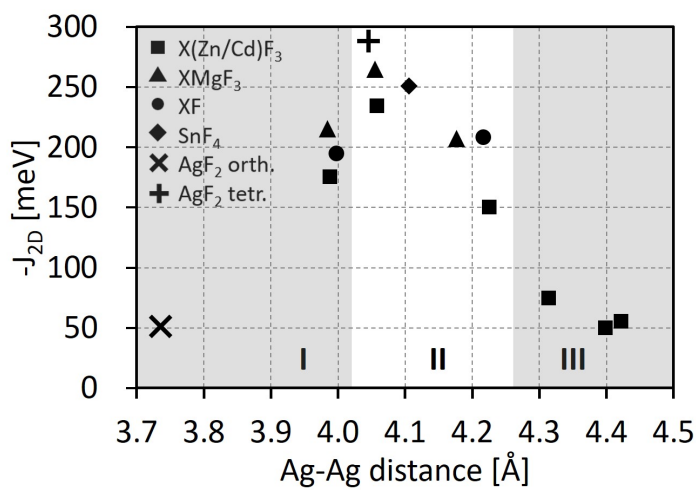

Figure 3. Dependence of antiferromagnetic coupling constant (–$J_{2D}$) on Ag-Ag distance – on different surfaces and in bulk AgF$_2$. Roman numerals and shading indicate three key domains (see text). $J_{2D}$ for bulk AgF$_2$, as calculated using PBEsol+U, are shown for both ground state orthorhombic and a hypothetical tetragonal structure. "X" in formulas stands for alkali metal cation.

Flattening of the AgF$_2$ layers should naturally result in enhancement of the AFM interactions within the sheets. Indeed, an increase in the intensity of superexchange interactions in comparison to bulk AgF$_2$ is readily apparent in domains I and II (table 1). The highest value of $J_{2D}$ (ca. –265 meV) was obtained for RbMgF$_3$ substrate, i.e. one with the shortest Ag-Ag distance within domain II. The $J_{2D}$ calculated for bulk AgF$_2$ (with corrugated layers) using the same methodology is over 5-fold smaller (–51 meV). For the promising RbMgF$_3$ system, we additionally calculated $J_{2D}$ using HSE06 hybrid functional, [30] which yielded values of –265.2 meV – virtually identical to the PBEsol+U value. $J_{2D}$ is also enhanced in domain I, although to a lesser extent because of buckling of Ag-F-Ag bond, which is still substantial (150-160°), yet less pronounced than in bulk AgF$_2$ (129°).

Errors in the computation of J can be estimated comparing with experimental values. Our PBEsol+U estimate for bulk AgF$_2$, $J_{2D}$ = –50 meV is smaller than the experimental [1] value –70 meV which suggest that real values may be even larger, but our experience with the HSE06 functional is that it overestimates the experiment by ca. 10% in similar systems with strong AFM superexchange, such as La$_2$CuO$_4$ or CsAgF$_3$. [37] In that case, the normalized value for $J_{2D}$ in RbMgF$_3$-AgF$_2$ surface system would be equal to ca. –240 meV. At any rate, both PBEsol+U and HSE06 results for some of our systems surpass the values in several cuprates (100–150 meV) roughly by a factor of 2.

As table 1 and fig. 3 show, $J_{2D}$ value drops steadily with Ag-Ag distance in domain II, as could be expected, and it is substantially decreased in domain III. Since substrates in the latter range have a significantly larger lattice constant, and Ag-F-Ag superexchange pathway involves short and long Ag-F separations, a much smaller $J_{2D}$ in this system is unsurprising. Notice that a different type of antiferrodistortive Ag-F bond ordering leads to ferromagnetic interactions in AgF$_2$ layers in fluoroargentates with double perovskite crystal structure. [5] In that case, a lobe of an active $d(x^2-y^2)$ orbital is nearly orthogonal to the basal plane of the neighbor $d(x^2-y^2)$ orbitals so that the effective hybridization tends to cancel. This is not the case here where the active orbitals have $d(z^2)$ symmetry in domain III so that nearest neighbor active d-orbitals hybridize with the same bridging p-orbital (see ESI for a modeling of this effect).

We were interested to see whether there exists a simple relationship between the strength of AFM interactions and the Ag-Ag distance in the monolayer (fig. 3). As expected, $J_{2D}$ appears to gradually decrease with the increasing distance in the studied range, but remains larger than typical cuprates (greater than 150 meV up to *ca.* 4.2 Å distance). A strong decrease, down to only tens of meV is found in **domain III** due to the antiferro orbital ordering. These changes can be attributed to the modulation of the hybridization matrix elements among active-d and p orbitals in Ag and F both due to changes in bond length, R, and Ag-F-Ag angle, $\alpha$. A perturbative computation yields the dominant superexchange contribution behaving as $J_{2D} \propto \cos^2\alpha / R^{16}$. As shown in the ESI, this expression captures qualitatively and even semiquantitative the main trends in the material dependence of magnetic interactions.

Analysis of the three-dimensional magnetization show that magnetic interactions are confined to the AgF$_2$ monolayer (see SI Fig. S2.1 for the example case of RbMgF$_3$ substrate). Again, this makes the studied systems very similar to undoped oxocuprates, where strong magnetic interactions are confined to [CuO$_2$]$^{2-}$ layers, both for the bulk systems [38] as well as for single layer ones. [39]

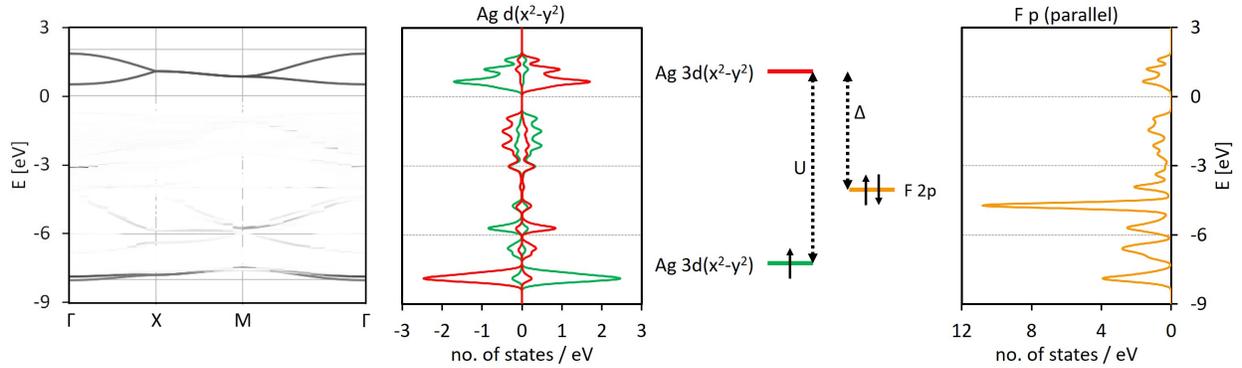

Fig. 4. Orbital-projected eDOS (Ag d($x^2$–$y^2$) and F p [parallel to Ag-F-Ag bonds] contributions) and band structure of AgF$_2$ layer on RbMgF$_3$ surface. Only the LHB, UHB and valence band are shown. The positions of level are schematic for clarity of presentation.

In order to gain insight into electronic and magnetic properties of AgF$_2$ monolayer on fluoride substrate, we also calculated the electronic band structure and density of states (eDOS) for all systems, an example of which is shown in fig. 4. The overall appearance is similar to bulk AgF$_2$, [1] *i.e.* that of a charge-transfer insulator, with lower Hubbard band (LHB) at *ca.* –8 eV and upper Hubbard band (UHB) at *ca.* 2 eV, while the ligand band is located between the two and centered at *ca.* –4 eV. There is strong admixing between Ag and F states just below the Fermi level, which indicates a marked covalence of Ag–F bonds. [40] The band gap amounts to *ca.* 1 eV – less than the 1.4 eV value for the bulk AgF$_2$ (calculated with the same method), which helps to understand why a flat system in the absence of a rigid fluoride support has a tendency to pucker.

In addition to all fluoride systems mentioned above, we studied AgF$_2$ deposited on an oxide surface – MgO. Similar systems involving superlattices of ternary silver(II) fluorides and ternary titanium oxides have previously been studied. [41,42] However, there are only a few known examples of compounds in which Ag(II) atoms are coordinated by oxygen, the most important being AgSO$_4$ [43] and several fluorosulfates, e.g. Ag(SO$_3$F)$_2$. [44] In fact, due to the strong electronegativity of Ag(II) cation, all known Ag(II)-O systems can be seen as negative charge transfer insulators (similar to some nickelates [45]). Thus, all known Ag(II)-O systems are unstable towards charge transfer (self-doping) or charge density wave (mixed valence). [21]

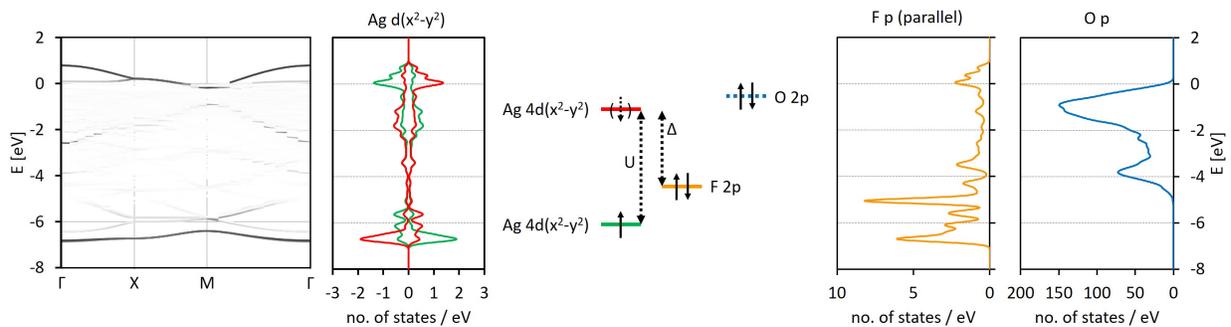

Fig. 5. Comparison of orbital-projected eDOS (Ag d($x^2$–$y^2$), F p [parallel to Ag-F-Ag bonds] and O 2p contributions) and band structure of AgF$_2$ layer on MgO surface. Dotted arrow in brackets on the upper 4d($x^2$-$y^2$) level indicates

that it is partially occupied due to charge transfer from O atoms. Only the LHB, UHB and valence band are shown. The position of levels is schematic for clarity of presentation.

Fig. 5 shows eDOS graphs for AgF$_2$ monolayer on MgO surface, together with eDOS for O atoms in the substrate. Although the overall structure of the Ag d($x^2$-$y^2$) part is similar to fig. 4, the position of the UHB is shifted down to around the Fermi level and is therefore partially populated. On the other hand, the position of O p bands and their partial overlap with UHB of Ag shows that it is partially depopulated; the aforementioned charge transfer between electron-poor Ag(II) and electron-rich O$^{2-}$ anion takes place, resulting in an electron doped AgF$_2$ layer. Doping is evident from depletion of spin on Ag atoms; for example, in a typical fluoride substrate the layer adjacent to AgF$_2$ one carries only minor spin contamination (ca. 0.01 $\mu_B$ per adjacent F atom), with the AgF$_2$ layer carrying null spin (with ±0.49 $\mu_B$ on each Ag center); however, the spin transfer to the MgO subsurface is as large as 0.06 $\mu_B$, with the net spin of 0.18 $\mu_B$ on each Ag center of the AgF$_2$ layer. This implies that as much as 25% of spin is depleted from the AgF$_2$ layer by placing it on the MgO substrate. To quantify the degree of doping, we compared the integrated eDOS up to $E_{Fermi}$ (as a measure of electron count) of AgF$_2$ monolayer on a substrate and the exact same monolayer (i.e. as optimized on a given substrate) isolated in vacuum. For the MgO system, we obtained the value of 0.34 excess electron per Ag atom, which is in reasonable agreement with the aforementioned distribution of spin. Comparing to the phase diagram of cuprates this corresponds to a strongly overdoped system, so superconductivity is not expected in this case. For the RbMgF$_3$ system, we obtained a value of 0.06, which suggests that even on fluoride substrate some additional charge is transferred into AgF$_2$ monolayer. This, however, occurs through hybridization without adding carriers to the conduction band.

Analysis of spin density map of AgF$_2$ layer on MgO surface (cf. SI fig. S2.2) shows that, compared to fluoride substrates, there is much more spin density on Ag d($z^2$) orbitals. Conversely, non-zero spin density on O p(z) orbitals indicates substantial depopulation of O states as could be expected based on previous experimental research. [46] Moreover, the AFM state of AgF$_2$ monolayer on MgO becomes even less stable than the FM state, hence the positive (ferromagnetic) $J_{2D}$ in table 1. Notice that in this case, since the system is metallic, the interpretation of $J_{2D}$ as the interaction between localized moments loses its meaning. This reversal of the sign of the magnetic interaction suggest that doping the AgF$_2$ layer should destroy long-range antiferromagnetic order if quantum fluctuations were taken into account beyond the present DFT method, just as for cuprates. Interestingly, ferromagnetism has recently being observed for a strongly hole-overdoped cuprate, [47] similar to our theoretical finding in the case of an electron-doped cuprate analogue.

The geometry of the AgF$_2$ layer on MgO is distorted compared to fluoride substrates of similar unit cell size (table 1): the Ag-F-Ag bridge diverges from straight angle by over 12˚ due to the fact that F atoms are bound strongly by Mg(II) cations from the surface. This suggests the presence of additional strain within the monolayer, as the same phenomenon was observed for flat layers optimized on substrates in **domain I**. Overall, the spin density map and eDOS graphs in fig. 5, together with marked distortions in the monolayer suggest that deposition of AgF$_2$ layer on oxide surfaces will likely result in destructive charge-transfer which also affects magnetism in undesirable way. Similar conclusions were obtained in the aforementioned computational studies of superlattices, where the AFM state was destabilized due to electronic reconstruction between AgF$_2$ and TiO$_2$ layers. [41,42]

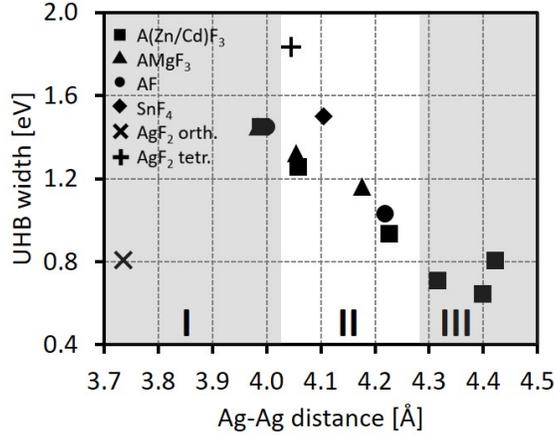

Fig. 6. Dependence of upper Hubbard band (UHB) width on Ag-Ag distance in tetragonal AgF$_2$ monolayer. Roman numerals and shading indicate three key domains (see text). Values for bulk AgF$_2$ as calculated using PBEsol+U, are shown for both ground state orthorhombic (X) and a hypothetical tetragonal structure (+). "A" in formulas stands for alkali metal cation.

In order to explore the possibility to metalize the system by electron doping, in fig. 6 we plotted the dependence of width of upper Hubbard band in AgF$_2$ monolayer on separation of Ag atoms within the monolayer. The general trend is similar as for $J_{2D}$ since the processes for an electron to hop to an equivalent site are similar to the ones responsible for superexchange. Thus, the bandwidth increases with the reduction of interatomic distance, which is expected, given the decrease of overlap between atomic orbitals with distances as discussed in detail in the ESI. Widening of the UHB band in the compressed monolayers is a desired result, as it makes self-trapping (polaron formation) of additional charges less likely and facilitates metallization by electron doping of AgF$_2$ monolayer. This effect adds to the reduction of linear coupling with out-of-plane ligand modes mentioned in the introduction.[2]

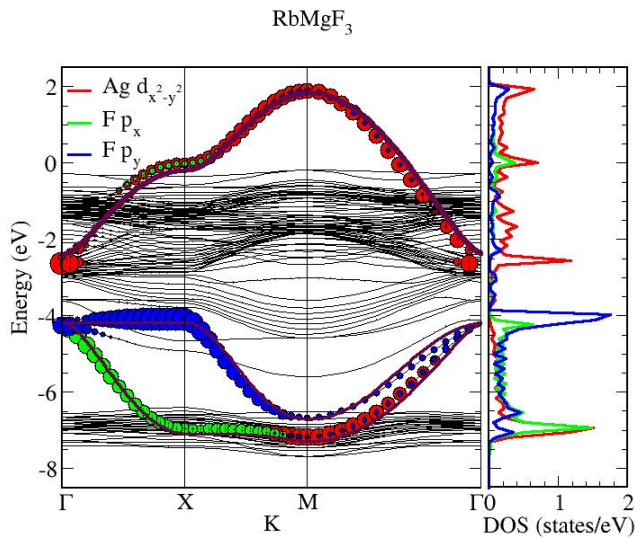

Fig. 7. Band structure of AgF$_2$ monolayer on an RbMgF$_3$ substrate. Circles show the character of the bands in Ag d(x$^2$-y$^2$) orbitals (red) and in F p(x) and p(y) orbitals oriented parallel to the Ag-F bond (green and blue). The right panel shows the density of states projected on the same orbitals. Thick brown lines show the fit with a three-band tight binding model.

In Fig. 7 we show the band structure for RbMgF$_3$ in the nonmagnetic (metallic) solution. Most of the bands originate in the substrate (thin lines). The circles show the character of the bands in the Ag d(x$^2$-y$^2$) orbitals and in the F p(x) and p(y) orbitals parallel to the bonds. The band structure projected on these orbitals is remarkably similar to high-temperature superconducting cuprates band structure projected on analogous orbitals of Cu and O. [48–50] The same can be said for the density of states shown in the right panel. The similarity can be further quantified by fitting the band structure with the three-band Hubbard model [49]. This model provides an excellent description of the relevant bands (fig. 7, and ESI Fig. S4.1). As can be seen from obtained values (Table 2), the $t_{pd}$ hopping parameter is comparable or even slightly larger than the corresponding parameter for cuprates (1.3–1.6 eV), [48,50] while the $t_{pp}$ hopping parameter is somewhat smaller than cuprates (0.6–0.65 eV). Overall the bandwidth of the conduction band is very similar to cuprates which will be used below.

A decrease in Ag-Ag distance from 4.06 Å in RbMgF$_3$ to 3.99 Å in KMgF$_3$ leads to a larger value of $t_{pd}$, but it is important to remember that this may decreases the effective hybridization as discussed in Ref. 1. In the case of KMgF$_3$, we report in Table 1 and Fig. S4.1 both the ground state corrugated solution and a metastable tetragonal solution. Although the Ag-F-Ag angle is not small (159.3°), the fit with the three-band model is still quite good, in contrast with the case of bulk AgF$_2$ where a five-band model is needed due to the much larger bending angle (132.4°) promoting strong mixing with an additional p-orbital.[1] Indeed, the band structure of the monolayer orthorhombic solution does not differ much from the tetragonal solution and the renormalization of the hopping matrix elements is modest (c.f. Table 2 and Fig. S4.1). This suggest that also these systems may mimic cuprates providing an interesting range of electronic properties to study the effects on a possible superconducting state.

Table 2. Hopping parameters from the Hubbard model for AgF$_2$ monolayer on selected substrates. *Values for bulk AgF$_2$ are taken from ref. [1] [1]. In this case we report also the hopping matrix element with an additional p-orbital perpendicular to the bond (in parenthesis) which needs to be considered for strong buckling. We report the nearest neighbor hoping between Ag d$_{x2-y2}$ and F p orbitals oriented along the bond ($t_{pd}$), nearest neighbor p(x) and p(y) orbitals ($t_{pp}$) and p(x)-p(x) and p(y)-p(y) among next nearer neighbor F's bridged by and Ag ($t'_{pp}$).

|  | RbMgF$_3$ tetragonal (flat) | KMgF$_3$ tetragonal (flat) | KMgF$_3$ orthorhombic (corrugated) | *bulk AgF$_2$ orthorhombic (corrugated) |
|---|---|---|---|---|
| a$_{Ag-Ag}$ [Å] | 4.06 | 3.99 | 3.99 | 3.74 |
| $t_{pd}$ [eV] | 1.60 | 1.72 | 1.60 | 1.24 (0.65) |
| $t_{pp}$ [eV] | 0.46 | 0.48 | 0.44 | 0.13-0.30 |
| $t'_{pp}$ [eV] | 0.18 | 0.19 | 0.17 |  |

To quantify the possibilities of AgF$_2$ monolayers as high-$T_c$ superconductors we first reexamine the relation between $T_c$ and $J_{2D}$ in cuprates. Assuming a $tJ$-model description, the $T_c/t$ ratio should be a function of $\eta=-J_{2D}/t$, where $t$ is the one-band hopping. More precisely, for a two-dimensional material the superconducting ordering temperature should be determined by the superfluid stiffness, [51,52]

$$k_B T_c = C\, n_s(\eta)\, (-J_{2D}) \quad (1)$$

with $n_s$ the density of superfluid pairs and C a constant of order 1. Following Ref. [51] we are assuming that the effect of $J_{2D}$ is twofold. It provides the attraction to create bound fermion pairs [53–55] and it provides the matrix elements to make the pairs mobile. Also, within the $tJ$-model the dimensionless superfluid density at optimum doping should be a function of $\eta$. Since bound pairs require a minimum value of $\eta$ we expect that $n_s(\eta) \to 0$ below a critical ratio $\eta_c$. We therefore expand, $k_B T_c/t = C\, n_s(\eta)\eta = C\, n_s'(\eta_c)\, \eta_c\, (\eta-\eta_c)$. Following Moreira et al. [56] we restrict to bulk monolayer cuprates where layers are weakly coupled and eq (1) applies without the need of additional corrections due to interlayer couplings. The left panel of fig. 8 shows that indeed there is a strong linear correlation between the maximum $T_c/t$ of each monolayer cuprate family and $\eta$ (detailed data and references can be found in the ESI). We took the $t$ and $J_{2D}$ values from accurate quantum chemistry computations in Ref. [56] where a similar linear relation was found with the same value of $\eta_c$=0.2. Nicely, such value is very close to the critical value, above which bound pairs are believed to exist in the $tJ$-model. [53,55]

Notice that Moreira et al. considered the absolute value of $T_c$ in Kelvin which gives an almost as good correlation as the one for dimensionless quantities considered here. This is simply because the variation of $t$ among different families is much more modest than the variation of $J_{2D}$. Fitting a band with nearest neighbor hopping to the conduction band of the paramagnetic solution of the AgF$_2$ monolayer on an RbMgF$_3$ we find $t$=530meV again very close to typical cuprate values. [56] Thus, in first approximation, we can use common absolutes units for both monolayer cuprates and silver fluorides and extrapolate from the cuprate region to the AgF$_2$ domain. In addition, this allows to include cuprate materials for which quantum chemistry computations of $t$ are not available but the experimental $J_{2D}$ can be obtained from two-magnon Raman scattering. This is shown in the right panel of Fig. 7 where we plot the cuprate data (blue and orange points). Making a linear regression we obtain a sloop of 1.2K/meV. A similar positive correlation between $T_c$ and $J_2$ has been found in strain engineered La$_2$CuO$_4$. [57] It should be noted that for multilayer materials the situations is more complicated with contrasting [52,57] results beyond our scope.

Using a linear extrapolation (dashed line) we obtain the expected values of the maximum $T_c$ attainable as a function of doping in each one of the AgF$_2$ monolayers (purple). We see that for the largest $J_{2D}$ monolayer-substrate combination (**RbMgF$_3$**), $T_c$ near 200K is achievable.

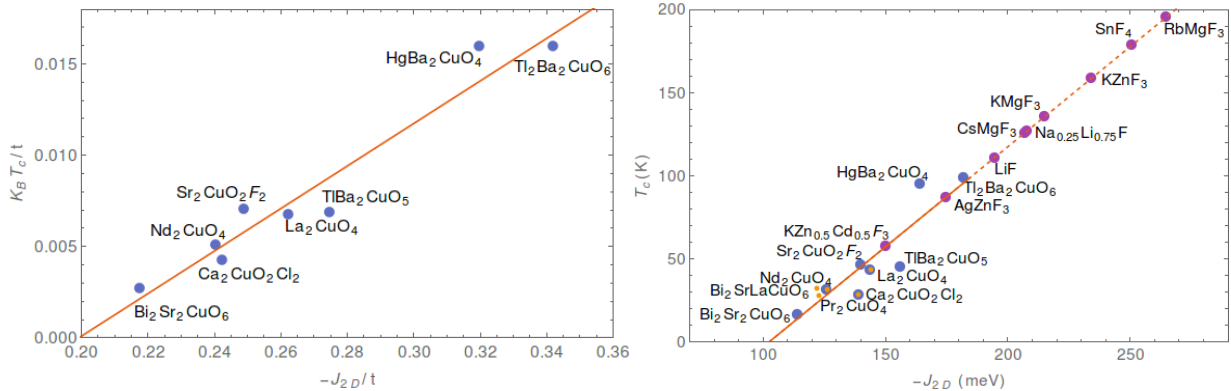

Fig. 8. Left panel: The maximum $T_c$ value of each cuprate family (in units of $t$) as a function of $J_{2D}$ in the same units. The line is the linear fit $k_B T_c/t = 0.12(\eta - \eta_c)$. Right panel: Blue and orange dots are the maximum experimental $T_c$ value in Kelvin vs $J_{2D}$ in meV for cuprate families containing one layer in the unit cell. Blue dots corresponds to the $J_{2D}$ computed in Ref. [56] while small orange dots are experimental $J_{2D}$ from two-magnon data. [58–62] Notice that for La$_2$CuO$_4$, Nd$_2$CuO$_4$ and Ca$_2$CuO$_2$Cl$_2$ both Raman and the quantum chemistry determination of $J_{2D}$ are available. The almost perfect overlap of the points serves as a test of the accuracy of the quantum chemistry computations. The line is the linear regression over the cuprate data with slope 1.2 K/meV. The purple dots are the expected values of the maximum $T_c$ of the AgF$_2$ monolayers considered in this work and labeled by the substrates. More detail on the data is reported in the ESI.

**Conclusions**

Goodenough-Kanamori-Anderson rules, [7] as well as recent computational work on AgF$_2$ [10] provide strong indications that AFM superexchange should be strongly enhanced in flat-layered AgF$_2$, in particular when the Ag-F-Ag angles will approach 180°. [1] Our computations indicate that a monolayer of silver(II) fluoride deposited on metal fluoride substrate has a potential for constituting a stable two-dimensional antiferromagnet, with superexchange interactions from 3- to over 5-fold enhanced as compared to bulk AgF$_2$ and simultaneously nearly 2-fold larger than La$_2$CuO$_4$.

There appears to be an optimal range of unit cell size of fluoride substrate, below which the deposited monolayer is prone to corrugation and above which it breaks down into molecules. Even for the corrugated monolayers the properties are much better than bulk AgF$_2$ with respect to the possibility to render the material metallic by doping and hopefully superconducting. In this respect, corrugation could be a helpful tool to study trends in electronic and superconducting properties as bandwidth is changed. For systems showing flat AgF$_2$ layers the width of the upper Hubbard band may reach 1.5 eV, which is nearly 2-fold larger than for bulk AgF$_2$ with corrugated layers.

We find that there are several compounds which can fulfill the very demanding chemical requirements imposed by AgF$_2$ and at the same time provide optimum heteroepitaxial conditions to create a flat monolayer. Our results clearly show that an AgF$_2$ layer with small buckling and not too large Ag-Ag distance exhibits oxocuprate-like electronic properties both in the metallic phase and in the insulating antiferromagnetic phase. The increased width of upper Hubbard band for flat layer systems as compared to bulk AgF$_2$ (up to 1.4 eV vs 0.4 eV) [1] will likely decrease the tendency of the system for polaron localization making metallization more feasible.

Extrapolating from cuprates we argue that the maximum critical temperature of a flat AgF$_2$ monolayer may reach nearly 200 K, which is much larger than record multilayer cuprates (133 K). [63] Thus, our results pave the road for a new family of quantum materials with very similar characteristics as high-$T_c$ cuprates (but in the absence of copper and oxygen) which could be study in a field effect transistor setup[Bollinger2011] to provide new insights on the fundamental and still mysterious physics of doped Mott insulators and high-$T_c$ superconductors.

**Acknowledgements**

WG thanks the Polish National Science Center (NCN) for the Maestro project (2017/26/A/ST5/00570). This research was carried out with the support of the Interdisciplinary Centre for Mathematical and Computational Modelling (ICM), University of Warsaw under grant ADVANCE++ (no. GA76-19). XPY acknowledges the support from the National Key Research and Development Program of China (Grants Numbers 2018YFA0305700, 2017YFA0403600), the National Natural Science Foundation of China (NSFC) (Grants Number 11674325), and support from the High Magnetic Field Laboratory of Anhui Province. JL acknowledges financial support from Italian MIUR through Project No. PRIN 2017Z8TS5B, and from Regione Lazio (L. R. 13/08) through project SIMAP. The work done in Hong Kong was supported by HKUST grant (R9418).

# Supplementary Information

S1. Strain Dependence of Magnetic Interactions

The leading change in magnetic interactions on the strained monolayers can be understood with a simple model in which all the dependence is attributed to variations in the hopping matrix elements due to bond elongation and bond bending. As in Ref. [1], we consider an Ag-F-Ag bridge and compute the magnetic interaction in perturbation theory in $t_{pd}$, the hopping matrix element between active d and p orbitals in Ag and F respectively. The latter can be parameterized in terms of the Slater-Koster [2] matrix element $pd\sigma$ and depends on the orbitals involved. In the case of a straight bond and a p-orbital pointing towards the lob of a $x^2 - y^2$ orbital, $t_{pd} = \sqrt{3}pd\sigma/2$. For the antiferro-orbital order depicted in Fig. 2 of the main text and relevant for domain (III) two other orbitals arrangements are needed: a p-orbital coaxial with a $z^2$-orbital for short $(s)$ bonds yielding $t_{pd(s)} = pd\sigma(s)$, and p- and $z^2$-orbitals with the axis perpendicular to each other for long $(l)$ bonds yielding $t_{pd(l)} = pd\sigma(l)/2$. Here, $(s)$ and $(l)$ stand for short and long bond.

We estimated the material and strain dependence using the data of Table 1 of the main manuscript and assuming the scaling of Ref. [3], namely the Slater-Koster matrix element scales as $pd\sigma \propto 1/R^4$ with $R$ the Ag-F bond length. This scaling was explicitly checked in Supplement of Ref. [4] comparing different DFT computations of $t_{pd}$. In case of bended bonds, the dependence on the Ag-F-Ag angle $\alpha$, is also given by the Slater-Koster expressions.

The equations for the exchange interaction are essentially the same as in Ref. [1], except that in region III where the bridge has a short and a long bond, the nearest neighbor hopping matrix element has to be substituted by its average, namely $t_{pd} = [t_{pd(s)} + t_{pd(l)}]/2 = pd\sigma(s)/2 + pd\sigma(l)/4$. We use the same parameters and conventions as in this paper unless otherwise specified. The magnetic interaction reads,

$$J = J^{(2)} + J^{(4,SE)} + J^{(4,HR)},$$

with a contribution due to direct pd ferromagnetic exchange ($K_{\parallel d} > 0$),

$$J^{(2)} = 2t_{pd}^2 \sin^2\left(\frac{\alpha}{2}\right)\left[\frac{1}{(\Delta - K_{\parallel d})} - \frac{1}{(\Delta + K_{\parallel d})}\right],$$

a dominant superexchange antiferromagnetic contribution,

$$J^{(4,SE)} = -t_{pd}^4 \cos^2\alpha \frac{1}{\Delta^2}\left[\frac{4}{U_d} + \frac{8}{(2\Delta + U_p)}\right],$$

and a small ferromagnetic contribution due to Hund's rule exchange interaction $J_H > 0$ on fluorine,

$$J^{(4,HR)} = t_{pd}^4 \sin^2\alpha \frac{4}{\Delta^2}\left[\frac{1}{(2\Delta + U_p - J_H)} - \frac{1}{(2\Delta + U_p)}\right].$$

Here factor of 2 typos in the expressions for $J^{(2)}$ and $J^{(4,HR)}$ in Ref. [1] have been corrected. Those typos do not affect the curves in Ref. [1].

We take a minimal set of parameters to illustrate the main trends: $K_{\|d} = 0.07$eV, $J_H = 0.7$, $\Delta = 2.9$eV for the charge transfer energy, $U_d = 9.4$eV, $U_p = 4$eV for the Coulomb repulsion on Silver and Fluorine respectively and $t_{pd}^0 \equiv t_{pd}(R_0) = 1.1$eV for a reference Ag-F bond length $R_0 = 2.07$Å. The charge transfer energy is the only parameter differing from Ref. [1] where $\Delta = 2.7$ eV. This takes into account the different functional used in the present work as discussed below. (Note: Here factor of 2 typos in the expressions for J(4,HR) and J(2) in Ref. [1] have been corrected. Those typos do not affect the published curves.)

In fig. S1.1 we compare the perturbative expressions with the results obtained with the full DFT solution.

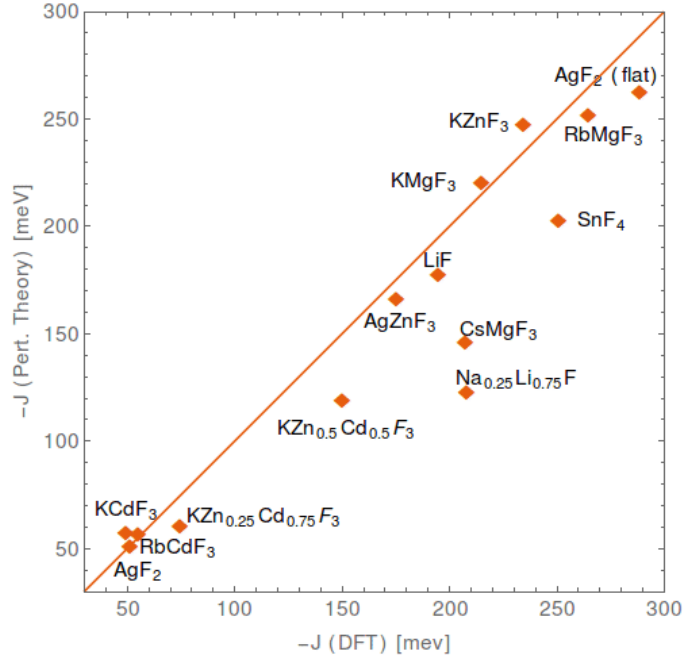

Figure S1.1: Magnetic interactions in AgF$_2$ monolayers obtained using perturbation theory vs. values obtained using DFT. Points are labelled by the substrate in which the monolayer is deposited. We also include the bulk form of AgF$_2$ and the hypothetical flat allotrope.

We see that the trends in the changes of magnetic interaction are in general well explained by the simple model. Assuming the dominant contribution is superexchange and according to the above equations one expects that the leading changes due to the structure are $J^{(4,SE)} \propto t_{pd}^4 \cos^2\alpha \propto \cos^2\alpha/R^{16}$. Figure S1.2 shows the magnetic interaction as a function of $1/R^{16}$. The straight lines show the superexchange contribution alone for different representative values of $\alpha$ and shifted by a constant to take approximately into account direct exchange, namely $-J^{(4,SE)} + 30$meV. The blue line corresponds to $\alpha = 180°$ (to be compared with the blue symbols corresponding to domains II and III), while the red full line corresponds to $\alpha = 154°$ which is near the value found in domain I (red circles). Finally, $\alpha = 129°$ corresponds to bulk AgF$_2$. Considering that the same parameter set is used in all cases and that the distance dependence of direct exchange is neglected the explanation of the main trends is quite satisfactory. That said, one should be aware that perturbation theory is only qualitatively valid for strongly covalent systems such as AgF$_2$ and cuprates. Indeed, we find that the perturbative expression overestimate the exact singlet triplet splitting of a flat Ag-F-Ag bridge at $R = R_0$=2.07 Å by 70% similar to what was found in cuprates. [5] We have absorbed such deficiency by taking the parameter $t_{pd}^0$ smaller than the more realistic estimates of Table 2 of the main text where $t_{pd}$ was estimated directly from the band structure.

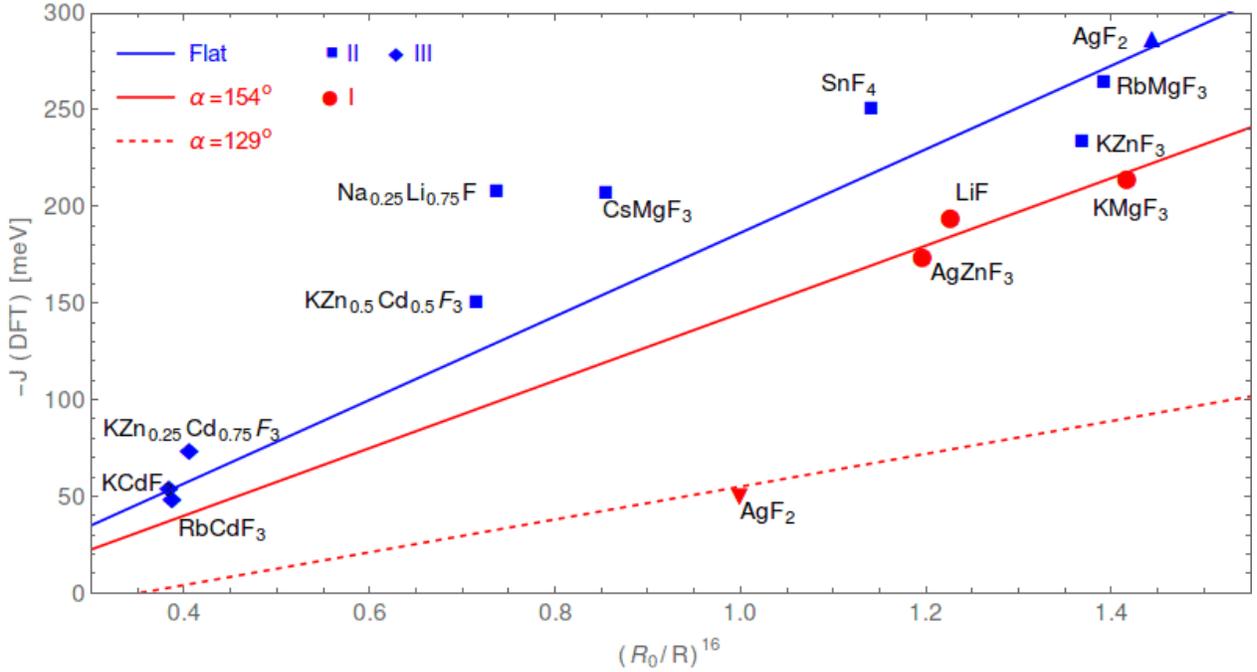

Figure S1.2: Structural dependence of magnetic interactions. J computed in DFT is plotted as a function of the inverse Ag-F distance to the 16th power for different values of the Ag-F-Ag angle. $\alpha = 129°$ and $R_0 = 2.07$ Å correspond to equilibrium AgF$_2$ (down pointing triangle ) flat AgF$_2$. The up pointing triangle is the hypothetical flat AgF$_2$ polymorph. In the case of antiferro-orbital ordering with a short $(R_{(s)})$ and a long bond $(R_{(l)})$, we defined an effective $R$ which absorbs the different overlap matrix element, $\sqrt{3}\,(1/R)^{1/4}/2 \equiv (1/R_{(s)})^{1/4}/2 + (1/R_{(l)})^{1/4}/4$.

Taking the same value of the charge transfer energy and Coulomb matrix element in all compounds is also a gross oversimplification as one expects a substantial material dependence. For example, a shorter Ag-apical F distance will tend to increase $\Delta$ and decrease $J$ in perturbation theory. Also, a reduction of screening by the substrate will tend to increase the Coulomb interactions and reduce $J$.

Another important methodological note is that the DFT+U and hybrid methods used here tend to yield values of $J$ smaller than the SCAN functional [6] used in Ref. [1] which, in turn, are in much better agreement with experiment as mentioned in the main text. For example for bulk AgF$_2$ one finds $J = -51$ meV (DFT+U), $J = -56$ meV (HSE06), $J = -71$ meV (SCAN, Ref. [1]), $J = -70$ meV (experiment, Ref. [4] ).

## S2. Spin density maps for selected systems.

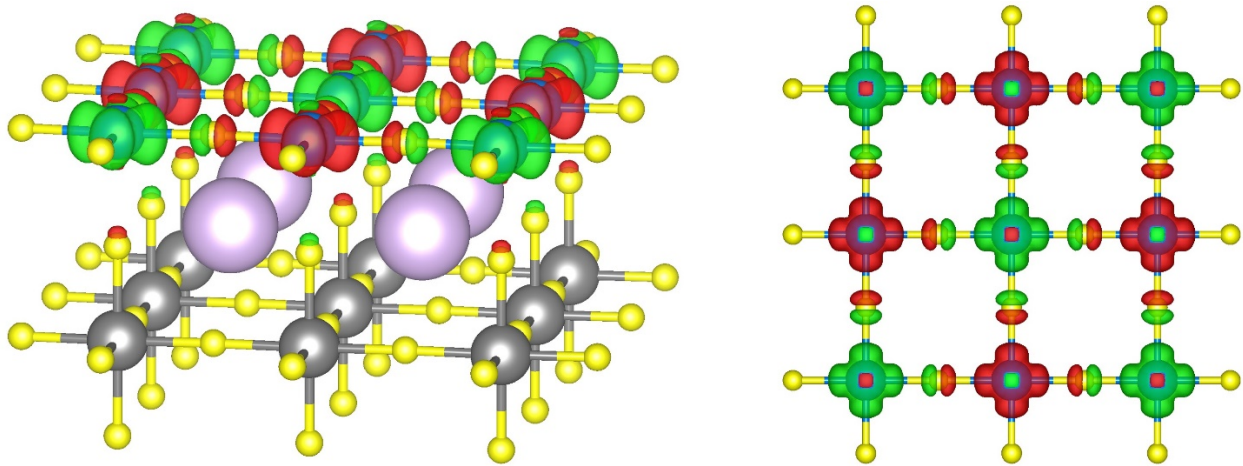

Fig. S2.1. Map of spin density in AgF$_2$ monolayer on RbMgF$_3$. Green – spin up, red – spin down. Atoms: Blue – Ag, yellow – F, grey – Mg, light purple - Rb.

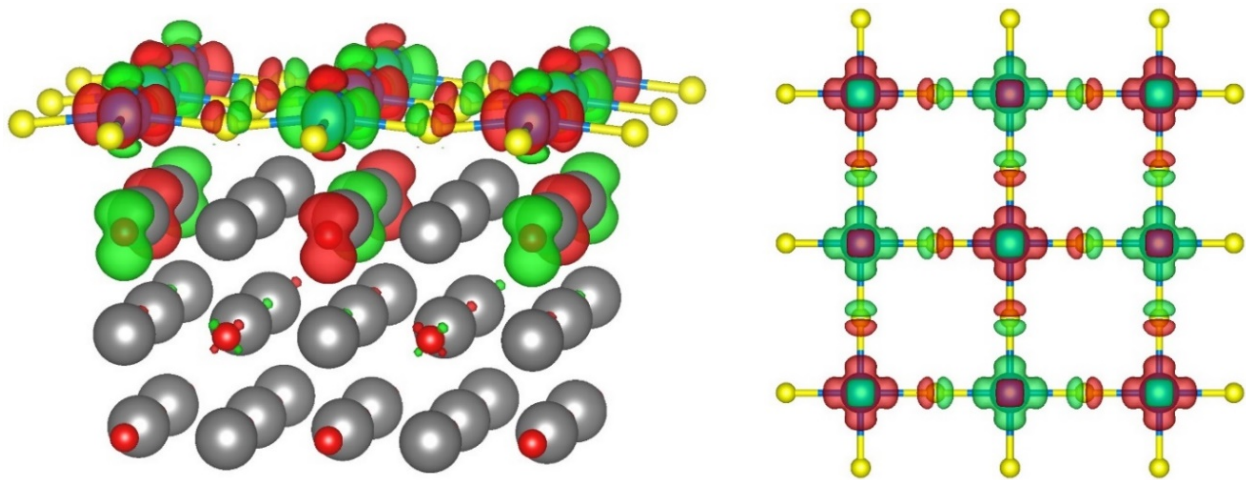

Fig. S2.2. Map of spin density in AgF$_2$ monolayer on MgO. Green – spin up, red – spin down. Atoms: Blue – Ag, yellow – F, grey – Mg, red – O.

S3. Electron localization function.

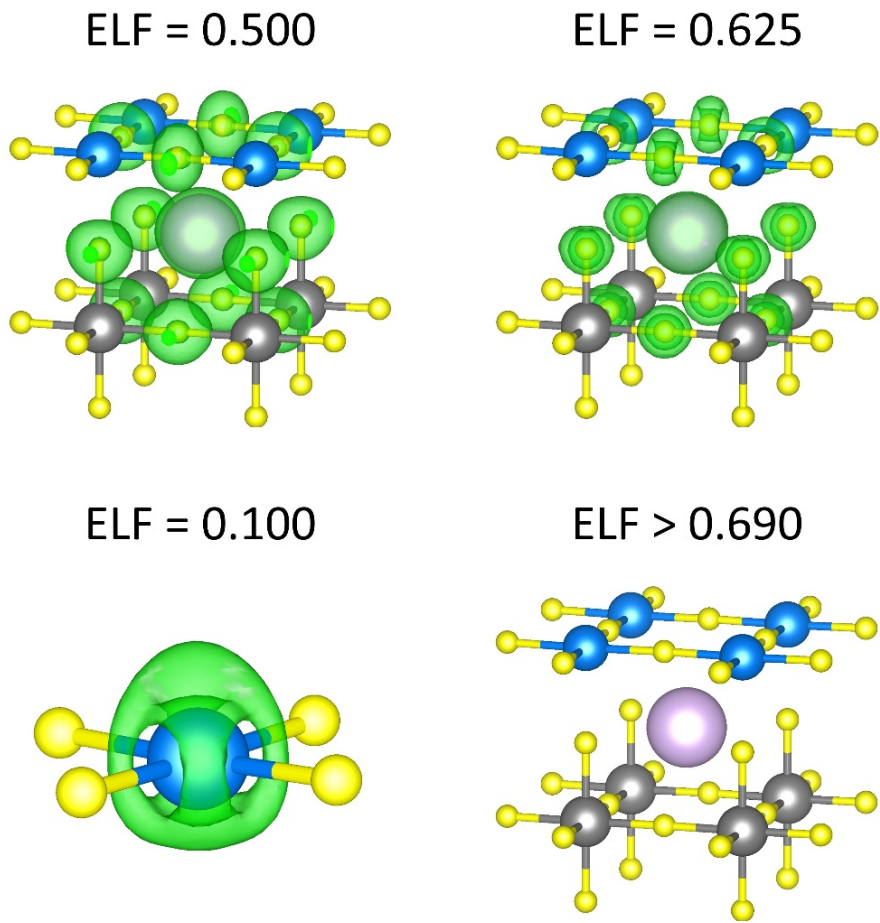

Fig. S3.1. Isosurfaces in AgF$_2$ monolayer and top layer of RbMgF$_3$ in AgF$_2$/RbMgF$_3$ system. For ELF = 0.100, only a fragment of the structure around one Ag atom is shown; isosurfaces around F atoms in that view are not shown for clarity.

The nature of bonding between AgF$_2$ monolayer and fluoride substrate can to some extent be inferred from electron localization function (ELF). [7] Fig. S3.1 shows isosurfaces for different values of ELF in the AgF$_2$ on RbMgF$_3$ system. The round shape of isosurfaces around F atoms in the Mg layer and around Rb atoms indicate ionic interactions within RbMgF$_3$ substrate, as is expected for this compound. The ring-shaped isosurfaces around F atoms in [AgF$_2$] layer correspond to lone pairs on their p(x)/p(y) orbitals. The valence attractor around Ag(II) atoms is flattened from the direction of top F atom of the substrate, as is the corresponding ELF structure around the said F atom. This indicates repulsion between Ag d(z$^2$) lone pair and F p lone pairs, and hints at ionic bonding between the two species. Therefore, AgF$_2$ monolayer appears to be bound to RbMgF$_3$ surface by largely ionic interactions between: a) Ag(II) and substrate F atoms b) F atoms in AgF$_2$ and Rb(I) atoms. ELF maps of AgF$_2$/RbMgF$_3$ interface (fig. S3.2) confirm these observations.

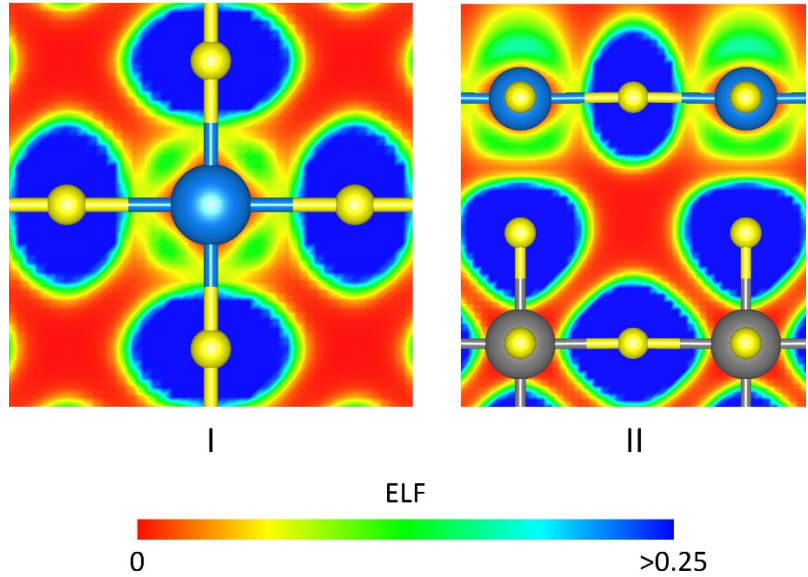

Fig. S3.2. ELF maps in AgF$_2$ monolayer on RbMgF$_3$. I – view of AgF$_2$ monolayer along *z* axis, II – view along *x/y* axis.

## S4. Electronic band structure for selected system, AgF$_2$ on KMgF$_3$ substrate, in ground state (corrugated) form as well as enforced flat geometry.

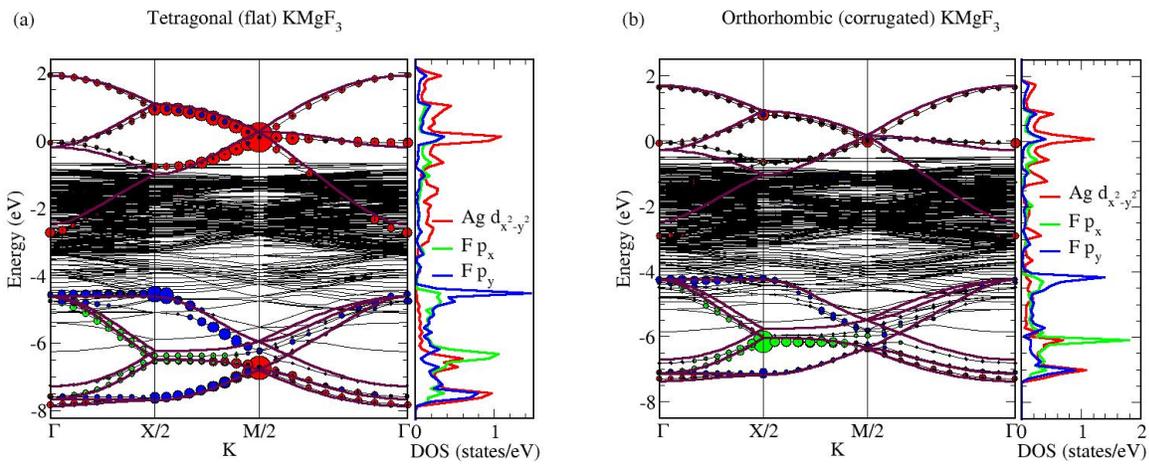

Fig. S4.1. Left panels: Band structures of AgF$_2$ monolayer on a KMgF$_3$ substrate in the case of the tetragonal solution (a) and orthorhombic solution (b). Circles show the character of the bands in Ag d$x^2$-$y^2$ orbitals (red) and in F p$_x$ and p$_y$ orbitals oriented parallel to the Ag-F bond (green and blue). The right panels show the density of states projected on the same orbitals. Thick brown lines on the left panels show the fits with the three-band tight binding model (see main text). Both solutions are shown in the orthorhombic unit cell for comparison. Results coming from spin-unpolarized calculation.

## S5. Cuprate data on monolayer materials.

We present the data used in Figure 8 of the main text. When possible we had used the midpoint $T_c$ from transport data which may be different to the onset $T_c$ often reported in the abstract.. Notice that cuprates have substantial corrections to the Heisenberg model so the reported $J$ should be intended as an effective parameter. We defined the Raman value as $J = \omega_R/2.8$ with $\omega_R$, the position of the maximum in the two-magnon Raman line shape.

| Material | $T_c$ max (K) | Ref. $T_c$ max | J Raman (mev) | Ref. Raman | J (meV) (ref. [8]) | t (meV) (ref. [8]) |
|---|---|---|---|---|---|---|
| La$_2$CuO$_4$ | 43 | [9] | 144 | [10] | 144 | 549 |
| Sr$_2$CuO$_2$F$_2$ | 46 | [10] | | | 140 | 562 |
| Ca$_2$CuO$_2$Cl$_2$ | 28 | [11] | 139 | [12] | 139 | 573 |
| TlBa$_2$CuO$_5$ | 45 | [13] | | | 156 | 568 |
| Tl$_2$Ba$_2$CuO$_6$ | 98.5 | [14] | | | 182 | 532 |
| HgBa$_2$CuO$_4$ | 95 | [15] | | | 164 | 513 |
| Bi$_2$Sr$_2$CuO$_6$ | 16.5 | [16] | | | 114 | 524 |
| Bi$_2$SrLaCuO$_6$ | 32 | [17] | 122 | [13] | | |
| Nd$_2$CuO$_4$ | 31 | [18] | 127 | [19] | 126 | 524 |
| Pr$_2$CuO$_4$ | 30 | [18] | 123 | [20] | | |

S5. Structures:

We investigated thirteen surface systems in this work in three different geometries of AgF$_2$ monolayer: flat (tetragonal), puckered (orthorhombic) and quasi-molecular. Not all systems were investigated in all three scenarios. Here we compile structures of all solutions relevant to the text:

**RbMgF₃ tetragonal**
```
  1.00000000000000
    8.1093997954999999    0.0000000000000000    0.0000000000000000
    0.0000000000000000    8.1093997954999999    0.0000000000000000
    0.0000000000000000    0.0000000000000000   48.3829994202000009
   Rb   Mg   F   Ag
   32   28   96   4
Direct
  0.2500000000000000  0.2500000000000000  0.1027422753542737
  0.2500000000000000  0.7500000000000000  0.1027422753542737
  0.7500000000000000  0.2500000000000000  0.1027422753542737
  0.7500000000000000  0.7500000000000000  0.1027422753542737
  0.2500000000000000  0.2500000000000000  0.1868236614785204
  0.2500000000000000  0.7500000000000000  0.1868236614785204
  0.7500000000000000  0.2500000000000000  0.1868236614785204
  0.7500000000000000  0.7500000000000000  0.1868236614785204
  0.2500000000000000  0.2500000000000000  0.2707796052825344
  0.2500000000000000  0.7500000000000000  0.2707796052825344
  0.7500000000000000  0.2500000000000000  0.2707796052825344
  0.7500000000000000  0.7500000000000000  0.2707796052825344
  0.2500000000000000  0.2500000000000000  0.3547599910000017
  0.2500000000000000  0.7500000000000000  0.3547599910000017
  0.7500000000000000  0.2500000000000000  0.3547599910000017
  0.7500000000000000  0.7500000000000000  0.3547599910000017
  0.2500000000000000  0.2500000000000000  0.4385600199999971
  0.2500000000000000  0.7500000000000000  0.4385600199999971
  0.7500000000000000  0.2500000000000000  0.4385600199999971
  0.7500000000000000  0.7500000000000000  0.4385600199999971
  0.2500000000000000  0.2500000000000000  0.5225744442254150
  0.2500000000000000  0.7500000000000000  0.5225744442254150
  0.7500000000000000  0.2500000000000000  0.5225744442254150
  0.7500000000000000  0.7500000000000000  0.5225744442254150
  0.2500000000000000  0.2500000000000000  0.6067066867392680
  0.2500000000000000  0.7500000000000000  0.6067066867392680
  0.7500000000000000  0.2500000000000000  0.6067066867392680
  0.7500000000000000  0.7500000000000000  0.6067066867392680
  0.2500000000000000  0.2500000000000000  0.6918917167010448
  0.2500000000000000  0.7500000000000000  0.6918917167010448
  0.7500000000000000  0.2500000000000000  0.6918917167010448
  0.7500000000000000  0.7500000000000000  0.6918917167010448
  0.0000000000000000  0.0000000000000000  0.1446678944334749
  0.0000000000000000  0.5000000000000000  0.1446678944334749
  0.5000000000000000  0.0000000000000000  0.1446678944334749
  0.5000000000000000  0.5000000000000000  0.1446678944334749
  0.0000000000000000  0.0000000000000000  0.2287638460475563
  0.0000000000000000  0.5000000000000000  0.2287638460475563
  0.5000000000000000  0.0000000000000000  0.2287638460475563
  0.5000000000000000  0.5000000000000000  0.2287638460475563
  0.0000000000000000  0.0000000000000000  0.3127690049650909
  0.0000000000000000  0.5000000000000000  0.3127690049650909
```

```
0.5000000000000000 0.0000000000000000 0.3127690049650909
0.5000000000000000 0.5000000000000000 0.3127690049650909
0.0000000000000000 0.0000000000000000 0.3966600050000011
0.0000000000000000 0.5000000000000000 0.3966600050000011
0.5000000000000000 0.0000000000000000 0.3966600050000011
0.5000000000000000 0.5000000000000000 0.3966600050000011
0.0000000000000000 0.0000000000000000 0.4805515679479912
0.0000000000000000 0.5000000000000000 0.4805515679479912
0.5000000000000000 0.0000000000000000 0.4805515679479912
0.5000000000000000 0.5000000000000000 0.4805515679479912
0.0000000000000000 0.0000000000000000 0.5645896087955331
0.0000000000000000 0.5000000000000000 0.5645896087955331
0.5000000000000000 0.0000000000000000 0.5645896087955331
0.5000000000000000 0.5000000000000000 0.5645896087955331
0.0000000000000000 0.0000000000000000 0.6485461040727000
0.0000000000000000 0.5000000000000000 0.6485461040727000
0.5000000000000000 0.0000000000000000 0.6485461040727000
0.5000000000000000 0.5000000000000000 0.6485461040727000
0.0000000000000000 0.5000000000000000 0.1036673970022548
0.0000000000000000 0.0000000000000000 0.1036673970022548
0.5000000000000000 0.0000000000000000 0.1036673970022548
0.5000000000000000 0.5000000000000000 0.1036673970022548
0.5000000000000000 0.2500000000000000 0.1447333744261741
0.7500000000000000 0.5000000000000000 0.1447333744261741
0.0000000000000000 0.2500000000000000 0.1447333744261741
0.2500000000000000 0.0000000000000000 0.1447333744261741
0.5000000000000000 0.7500000000000000 0.1447333744261741
0.2500000000000000 0.5000000000000000 0.1447333744261741
0.7500000000000000 0.0000000000000000 0.1447333744261741
0.0000000000000000 0.7500000000000000 0.1447333744261741
0.5000000000000000 0.0000000000000000 0.1867757831063699
0.5000000000000000 0.5000000000000000 0.1867757831063699
0.0000000000000000 0.0000000000000000 0.1867757831063699
0.0000000000000000 0.5000000000000000 0.1867757831063699
0.2500000000000000 0.5000000000000000 0.2287710454990730
0.7500000000000000 0.5000000000000000 0.2287710454990730
0.2500000000000000 0.0000000000000000 0.2287710454990730
0.0000000000000000 0.7500000000000000 0.2287710454990730
0.7500000000000000 0.0000000000000000 0.2287710454990730
0.5000000000000000 0.2500000000000000 0.2287710454990730
0.5000000000000000 0.7500000000000000 0.2287710454990730
0.0000000000000000 0.2500000000000000 0.2287710454990730
0.0000000000000000 0.0000000000000000 0.2707735090089329
0.5000000000000000 0.0000000000000000 0.2707735090089329
0.0000000000000000 0.5000000000000000 0.2707735090089329
0.5000000000000000 0.5000000000000000 0.2707735090089329
0.0000000000000000 0.7500000000000000 0.3127730853707575
0.7500000000000000 0.5000000000000000 0.3127730853707575
0.2500000000000000 0.0000000000000000 0.3127730853707575
0.5000000000000000 0.2500000000000000 0.3127730853707575
```

```
0.2500000000000000 0.5000000000000000 0.3127730853707575
0.0000000000000000 0.2500000000000000 0.3127730853707575
0.7500000000000000 0.0000000000000000 0.3127730853707575
0.5000000000000000 0.7500000000000000 0.3127730853707575
0.0000000000000000 0.0000000000000000 0.3547599910000017
0.0000000000000000 0.5000000000000000 0.3547599910000017
0.5000000000000000 0.0000000000000000 0.3547599910000017
0.5000000000000000 0.5000000000000000 0.3547599910000017
0.0000000000000000 0.7500000000000000 0.3966600050000011
0.2500000000000000 0.5000000000000000 0.3966600050000011
0.0000000000000000 0.2500000000000000 0.3966600050000011
0.7500000000000000 0.5000000000000000 0.3966600050000011
0.2500000000000000 0.0000000000000000 0.3966600050000011
0.7500000000000000 0.0000000000000000 0.3966600050000011
0.5000000000000000 0.2500000000000000 0.3966600050000011
0.5000000000000000 0.7500000000000000 0.3966600050000011
0.5000000000000000 0.0000000000000000 0.4385600199999971
0.0000000000000000 0.0000000000000000 0.4385600199999971
0.0000000000000000 0.5000000000000000 0.4385600199999971
0.5000000000000000 0.5000000000000000 0.4385600199999971
0.5000000000000000 0.2500000000000000 0.4806151204640515
0.2500000000000000 0.0000000000000000 0.4806151204640515
0.7500000000000000 0.5000000000000000 0.4806151204640515
0.0000000000000000 0.7500000000000000 0.4806151204640515
0.5000000000000000 0.7500000000000000 0.4806151204640515
0.7500000000000000 0.0000000000000000 0.4806151204640515
0.0000000000000000 0.2500000000000000 0.4806151204640515
0.2500000000000000 0.5000000000000000 0.4806151204640515
0.5000000000000000 0.0000000000000000 0.5226217637011789
0.0000000000000000 0.0000000000000000 0.5226217637011789
0.5000000000000000 0.5000000000000000 0.5226217637011789
0.0000000000000000 0.5000000000000000 0.5226217637011789
0.7500000000000000 0.0000000000000000 0.5646648787294885
0.2500000000000000 0.5000000000000000 0.5646648787294885
0.5000000000000000 0.2500000000000000 0.5646648787294885
0.2500000000000000 0.0000000000000000 0.5646648787294885
0.7500000000000000 0.5000000000000000 0.5646648787294885
0.0000000000000000 0.2500000000000000 0.5646648787294885
0.5000000000000000 0.7500000000000000 0.5646648787294885
0.0000000000000000 0.7500000000000000 0.5646648787294885
0.5000000000000000 0.5000000000000000 0.6066330134538273
0.0000000000000000 0.0000000000000000 0.6066330134538273
0.5000000000000000 0.0000000000000000 0.6066330134538273
0.0000000000000000 0.5000000000000000 0.6066330134538273
0.7500000000000000 0.5000000000000000 0.6488862154410254
0.5000000000000000 0.2500000000000000 0.6488862154410254
0.2500000000000000 0.5000000000000000 0.6488862154410254
0.0000000000000000 0.2500000000000000 0.6488862154410254
0.7500000000000000 0.0000000000000000 0.6488862154410254
0.5000000000000000 0.7500000000000000 0.6488862154410254
```

```
0.2500000000000000  0.0000000000000000  0.6488862154410254
0.0000000000000000  0.7500000000000000  0.6488862154410254
0.5000000000000000  0.5000000000000000  0.6899087424630737
0.0000000000000000  0.0000000000000000  0.6899087424630737
0.5000000000000000  0.0000000000000000  0.6899087424630737
0.0000000000000000  0.5000000000000000  0.6899087424630737
0.2500000000000000  0.0000000000000000  0.7403686153847279
0.2500000000000000  0.5000000000000000  0.7403686153847279
0.7500000000000000  0.0000000000000000  0.7403686153847279
0.7500000000000000  0.5000000000000000  0.7403686153847279
0.0000000000000000  0.2500000000000000  0.7403686153847279
0.5000000000000000  0.2500000000000000  0.7403686153847279
0.0000000000000000  0.7500000000000000  0.7403686153847279
0.5000000000000000  0.7500000000000000  0.7403686153847279
0.5000000000000000  0.5000000000000000  0.7397951250263464
0.0000000000000000  0.0000000000000000  0.7397951250263464
0.5000000000000000  0.0000000000000000  0.7397951250263464
0.0000000000000000  0.5000000000000000  0.7397951250263464
```

**RbMgF₃ orthorhombic**
  1.00000000000000
    8.1093997954999999    0.0000000000000000    0.0000000000000000
    0.0000000000000000    8.1093997954999999    0.0000000000000000
    0.0000000000000000    0.0000000000000000   48.3829994202000009
   Rb  Mg  F  Ag
   32  28  96  4
Direct
  0.2500000000000000  0.2500000000000000  0.1027422753542737
  0.2500000000000000  0.7500000000000000  0.1027422753542737
  0.7500000000000000  0.2500000000000000  0.1027422753542737
  0.7500000000000000  0.7500000000000000  0.1027422753542737
  0.2500000000000000  0.2500000000000000  0.1868236614785204
  0.2500000000000000  0.7500000000000000  0.1868236614785204
  0.7500000000000000  0.2500000000000000  0.1868236614785204
  0.7500000000000000  0.7500000000000000  0.1868236614785204
  0.2500000000000000  0.2500000000000000  0.2707796052825344
  0.2500000000000000  0.7500000000000000  0.2707796052825344
  0.7500000000000000  0.2500000000000000  0.2707796052825344
  0.7500000000000000  0.7500000000000000  0.2707796052825344
  0.2500000000000000  0.2500000000000000  0.3547599910000017
  0.2500000000000000  0.7500000000000000  0.3547599910000017
  0.7500000000000000  0.2500000000000000  0.3547599910000017
  0.7500000000000000  0.7500000000000000  0.3547599910000017
  0.2500000000000000  0.2500000000000000  0.4385600199999971
  0.2500000000000000  0.7500000000000000  0.4385600199999971
  0.7500000000000000  0.2500000000000000  0.4385600199999971
  0.7500000000000000  0.7500000000000000  0.4385600199999971
  0.2500000000000000  0.2500000000000000  0.5225976950118241
  0.2500000000000000  0.7500000000000000  0.5225976950118241
  0.7500000000000000  0.2500000000000000  0.5225976950118241
  0.7500000000000000  0.7500000000000000  0.5225976950118241
  0.2500000000000000  0.2500000000000000  0.6067565208269571
  0.2500000000000000  0.7500000000000000  0.6067565208269571
  0.7500000000000000  0.2500000000000000  0.6067565208269571
  0.7500000000000000  0.7500000000000000  0.6067565208269571
  0.2500000000000000  0.2500000000000000  0.6919554868408264
  0.2500000000000000  0.7500000000000000  0.6919554868408264
  0.7500000000000000  0.2500000000000000  0.6919554868408264
  0.7500000000000000  0.7500000000000000  0.6919554868408264
  0.0000000000000000  0.0000000000000000  0.1446678944334749
  0.0000000000000000  0.5000000000000000  0.1446678944334749
  0.5000000000000000  0.0000000000000000  0.1446678944334749
  0.5000000000000000  0.5000000000000000  0.1446678944334749
  0.0000000000000000  0.0000000000000000  0.2287638460475563
  0.0000000000000000  0.5000000000000000  0.2287638460475563
  0.5000000000000000  0.0000000000000000  0.2287638460475563
  0.5000000000000000  0.5000000000000000  0.2287638460475563
  0.0000000000000000  0.0000000000000000  0.3127690049650909
  0.0000000000000000  0.5000000000000000  0.3127690049650909

```
0.5000000000000000  0.0000000000000000  0.3127690049650909
0.5000000000000000  0.5000000000000000  0.3127690049650909
0.0000000000000000  0.0000000000000000  0.3966600050000011
0.0000000000000000  0.5000000000000000  0.3966600050000011
0.5000000000000000  0.0000000000000000  0.3966600050000011
0.5000000000000000  0.5000000000000000  0.3966600050000011
0.0000000000000000  0.0000000000000000  0.4805616843276963
0.0000000000000000  0.5000000000000000  0.4805616843276963
0.5000000000000000  0.0000000000000000  0.4805616843276963
0.5000000000000000  0.5000000000000000  0.4805616843276963
0.0000000000000000  0.0000000000000000  0.5646260373527311
0.0000000000000000  0.5000000000000000  0.5646260373527311
0.5000000000000000  0.0000000000000000  0.5646260373527311
0.5000000000000000  0.5000000000000000  0.5646260373527311
0.0000000000000000  0.0000000000000000  0.6485979530936599
0.0000000000000000  0.5000000000000000  0.6485979530936599
0.5000000000000000  0.0000000000000000  0.6485979530936599
0.5000000000000000  0.5000000000000000  0.6485979530936599
0.0000000000000000  0.5000000000000000  0.1036673970022548
0.0000000000000000  0.0000000000000000  0.1036673970022548
0.5000000000000000  0.0000000000000000  0.1036673970022548
0.5000000000000000  0.5000000000000000  0.1036673970022548
0.5000000000000000  0.2500000000000000  0.1447333744261741
0.7500000000000000  0.5000000000000000  0.1447333744261741
0.0000000000000000  0.2500000000000000  0.1447333744261741
0.2500000000000000  0.0000000000000000  0.1447333744261741
0.5000000000000000  0.7500000000000000  0.1447333744261741
0.2500000000000000  0.5000000000000000  0.1447333744261741
0.7500000000000000  0.0000000000000000  0.1447333744261741
0.0000000000000000  0.7500000000000000  0.1447333744261741
0.5000000000000000  0.0000000000000000  0.1867757831063699
0.5000000000000000  0.5000000000000000  0.1867757831063699
0.0000000000000000  0.0000000000000000  0.1867757831063699
0.0000000000000000  0.5000000000000000  0.1867757831063699
0.2500000000000000  0.5000000000000000  0.2287710454990730
0.7500000000000000  0.5000000000000000  0.2287710454990730
0.2500000000000000  0.0000000000000000  0.2287710454990730
0.0000000000000000  0.7500000000000000  0.2287710454990730
0.7500000000000000  0.0000000000000000  0.2287710454990730
0.5000000000000000  0.2500000000000000  0.2287710454990730
0.5000000000000000  0.7500000000000000  0.2287710454990730
0.0000000000000000  0.2500000000000000  0.2287710454990730
0.0000000000000000  0.0000000000000000  0.2707735090089329
0.5000000000000000  0.0000000000000000  0.2707735090089329
0.0000000000000000  0.5000000000000000  0.2707735090089329
0.5000000000000000  0.5000000000000000  0.2707735090089329
0.0000000000000000  0.7500000000000000  0.3127730853707575
0.7500000000000000  0.5000000000000000  0.3127730853707575
0.2500000000000000  0.0000000000000000  0.3127730853707575
0.5000000000000000  0.2500000000000000  0.3127730853707575
```

```
0.2500000000000000  0.5000000000000000  0.3127730853707575
0.0000000000000000  0.2500000000000000  0.3127730853707575
0.7500000000000000  0.0000000000000000  0.3127730853707575
0.5000000000000000  0.7500000000000000  0.3127730853707575
0.0000000000000000  0.0000000000000000  0.3547599910000017
0.0000000000000000  0.5000000000000000  0.3547599910000017
0.5000000000000000  0.0000000000000000  0.3547599910000017
0.5000000000000000  0.5000000000000000  0.3547599910000017
0.0000000000000000  0.7500000000000000  0.3966600050000011
0.2500000000000000  0.5000000000000000  0.3966600050000011
0.0000000000000000  0.2500000000000000  0.3966600050000011
0.7500000000000000  0.5000000000000000  0.3966600050000011
0.2500000000000000  0.0000000000000000  0.3966600050000011
0.7500000000000000  0.0000000000000000  0.3966600050000011
0.5000000000000000  0.2500000000000000  0.3966600050000011
0.5000000000000000  0.7500000000000000  0.3966600050000011
0.5000000000000000  0.0000000000000000  0.4385600199999971
0.0000000000000000  0.0000000000000000  0.4385600199999971
0.0000000000000000  0.5000000000000000  0.4385600199999971
0.5000000000000000  0.5000000000000000  0.4385600199999971
0.5000000000000000  0.2500000000000000  0.4806232980605312
0.2500000000000000  0.0000000000000000  0.4806232980605312
0.7500000000000000  0.5000000000000000  0.4806232980605312
0.0000000000000000  0.7500000000000000  0.4806232980605312
0.5000000000000000  0.7500000000000000  0.4806232821020091
0.7500000000000000  0.0000000000000000  0.4806232821020091
0.0000000000000000  0.2500000000000000  0.4806232821020091
0.2500000000000000  0.5000000000000000  0.4806232821020091
0.5000000000000000  0.0000000000000000  0.5226439677185604
0.0000000000000000  0.0000000000000000  0.5226439677185604
0.5000000000000000  0.5000000000000000  0.5226439677185604
0.0000000000000000  0.5000000000000000  0.5226439677185604
0.7500000000000000  0.0000000000000000  0.5647047739184170
0.2500000000000000  0.5000000000000000  0.5647047739184170
0.5000000000000000  0.2500000000000000  0.5647057839895298
0.2500000000000000  0.0000000000000000  0.5647057839895298
0.7500000000000000  0.5000000000000000  0.5647057839895298
0.0000000000000000  0.2500000000000000  0.5647047739184170
0.5000000000000000  0.7500000000000000  0.5647047739184170
0.0000000000000000  0.7500000000000000  0.5647057839895298
0.5000000000000000  0.5000000000000000  0.6066757381614923
0.0000000000000000  0.0000000000000000  0.6066757381614923
0.5000000000000000  0.0000000000000000  0.6066757381614923
0.0000000000000000  0.5000000000000000  0.6066757381614923
0.7500000000000000  0.5000000000000000  0.6489429539452467
0.5000000000000000  0.2500000000000000  0.6489429539452467
0.2500000000000000  0.5000000000000000  0.6489421240046040
0.0000000000000000  0.2500000000000000  0.6489421240046040
0.7500000000000000  0.0000000000000000  0.6489421240046040
0.5000000000000000  0.7500000000000000  0.6489421240046040
```

```
0.2500000000000000  0.0000000000000000  0.6489429539452467
0.0000000000000000  0.7500000000000000  0.6489429539452467
0.5000000000000000  0.5000000000000000  0.6899557227987971
0.0000000000000000  0.0000000000000000  0.6899557227987971
0.5000000000000000  0.0000000000000000  0.6899557227987971
0.0000000000000000  0.5000000000000000  0.6899557227987971
0.2500000000000000  0.0000000000000000  0.7401713499675725
0.2500000000000000  0.5000000000000000  0.7406355473415535
0.7500000000000000  0.0000000000000000  0.7406355473415535
0.7500000000000000  0.5000000000000000  0.7401713499675725
0.0000000000000000  0.2500000000000000  0.7406355473415535
0.5000000000000000  0.2500000000000000  0.7401713499675725
0.0000000000000000  0.7500000000000000  0.7401713499675725
0.5000000000000000  0.7500000000000000  0.7406355473415535
0.5000000000000000  0.5000000000000000  0.7398300973268677
0.0000000000000000  0.0000000000000000  0.7398300973268677
0.5000000000000000  0.0000000000000000  0.7398300973268677
0.0000000000000000  0.5000000000000000  0.7398300973268677
```

**RbMgF₃ quasi-molecular**
  1.00000000000000
     8.1093997954999999    0.0000000000000000    0.0000000000000000
     0.0000000000000000    8.1093997954999999    0.0000000000000000
     0.0000000000000000    0.0000000000000000   48.3829994202000009
   Rb  Mg  F  Ag
   32  28  96  4
Direct
  0.2500000000000000  0.2500000000000000  0.1027422753542737
  0.2500000000000000  0.7500000000000000  0.1027422753542737
  0.7500000000000000  0.2500000000000000  0.1027422753542737
  0.7500000000000000  0.7500000000000000  0.1027422753542737
  0.2500000000000000  0.2500000000000000  0.1868236614785204
  0.2500000000000000  0.7500000000000000  0.1868236614785204
  0.7500000000000000  0.2500000000000000  0.1868236614785204
  0.7500000000000000  0.7500000000000000  0.1868236614785204
  0.2500000000000000  0.2500000000000000  0.2707796052825344
  0.2500000000000000  0.7500000000000000  0.2707796052825344
  0.7500000000000000  0.2500000000000000  0.2707796052825344
  0.7500000000000000  0.7500000000000000  0.2707796052825344
  0.2500000000000000  0.2500000000000000  0.3547599910000017
  0.2500000000000000  0.7500000000000000  0.3547599910000017
  0.7500000000000000  0.2500000000000000  0.3547599910000017
  0.7500000000000000  0.7500000000000000  0.3547599910000017
  0.2500000000000000  0.2500000000000000  0.4385600199999971
  0.2500000000000000  0.7500000000000000  0.4385600199999971
  0.7500000000000000  0.2500000000000000  0.4385600199999971
  0.7500000000000000  0.7500000000000000  0.4385600199999971
  0.2500000000000000  0.2500000000000000  0.5225750986204483
  0.2500000000000000  0.7500000000000000  0.5225750986204483
  0.7500000000000000  0.2500000000000000  0.5225750986204483
  0.7500000000000000  0.7500000000000000  0.5225750986204483
  0.2500000000000000  0.2500000000000000  0.6067078538161043
  0.2500000000000000  0.7500000000000000  0.6067078538161043
  0.7500000000000000  0.2500000000000000  0.6067078538161043
  0.7500000000000000  0.7500000000000000  0.6067078538161043
  0.2500000000000000  0.2500000000000000  0.6919185109015361
  0.2500000000000000  0.7500000000000000  0.6919185109015361
  0.7500000000000000  0.2500000000000000  0.6919185109015361
  0.7500000000000000  0.7500000000000000  0.6919185109015361
  0.0000000000000000  0.0000000000000000  0.1446678944334749
  0.0000000000000000  0.5000000000000000  0.1446678944334749
  0.5000000000000000  0.0000000000000000  0.1446678944334749
  0.5000000000000000  0.5000000000000000  0.1446678944334749
  0.0000000000000000  0.0000000000000000  0.2287638460475563
  0.0000000000000000  0.5000000000000000  0.2287638460475563
  0.5000000000000000  0.0000000000000000  0.2287638460475563
  0.5000000000000000  0.5000000000000000  0.2287638460475563
  0.0000000000000000  0.0000000000000000  0.3127690049650909
  0.0000000000000000  0.5000000000000000  0.3127690049650909

```
0.5000000000000000 0.0000000000000000 0.3127690049650909
0.5000000000000000 0.5000000000000000 0.3127690049650909
0.0000000000000000 0.0000000000000000 0.3966600050000011
0.0000000000000000 0.5000000000000000 0.3966600050000011
0.5000000000000000 0.0000000000000000 0.3966600050000011
0.5000000000000000 0.5000000000000000 0.3966600050000011
0.0000000000000000 0.0000000000000000 0.4805551818670854
0.0000000000000000 0.5000000000000000 0.4805551818670854
0.5000000000000000 0.0000000000000000 0.4805551818670854
0.5000000000000000 0.5000000000000000 0.4805551818670854
0.0000000000000000 0.0000000000000000 0.5645868816716905
0.0000000000000000 0.5000000000000000 0.5645868816716905
0.5000000000000000 0.0000000000000000 0.5645868816716905
0.5000000000000000 0.5000000000000000 0.5645868816716905
0.0000000000000000 0.0000000000000000 0.6485165036821239
0.0000000000000000 0.5000000000000000 0.6485165036821239
0.5000000000000000 0.0000000000000000 0.6485165036821239
0.5000000000000000 0.5000000000000000 0.6485165036821239
0.0000000000000000 0.5000000000000000 0.1036673970022548
0.0000000000000000 0.0000000000000000 0.1036673970022548
0.5000000000000000 0.0000000000000000 0.1036673970022548
0.5000000000000000 0.5000000000000000 0.1036673970022548
0.5000000000000000 0.2500000000000000 0.1447333744261741
0.7500000000000000 0.5000000000000000 0.1447333744261741
0.0000000000000000 0.2500000000000000 0.1447333744261741
0.2500000000000000 0.0000000000000000 0.1447333744261741
0.5000000000000000 0.7500000000000000 0.1447333744261741
0.2500000000000000 0.5000000000000000 0.1447333744261741
0.7500000000000000 0.0000000000000000 0.1447333744261741
0.0000000000000000 0.7500000000000000 0.1447333744261741
0.5000000000000000 0.0000000000000000 0.1867757831063699
0.5000000000000000 0.5000000000000000 0.1867757831063699
0.0000000000000000 0.0000000000000000 0.1867757831063699
0.0000000000000000 0.5000000000000000 0.1867757831063699
0.2500000000000000 0.5000000000000000 0.2287710454990730
0.7500000000000000 0.5000000000000000 0.2287710454990730
0.2500000000000000 0.0000000000000000 0.2287710454990730
0.0000000000000000 0.7500000000000000 0.2287710454990730
0.7500000000000000 0.0000000000000000 0.2287710454990730
0.5000000000000000 0.2500000000000000 0.2287710454990730
0.5000000000000000 0.7500000000000000 0.2287710454990730
0.0000000000000000 0.2500000000000000 0.2287710454990730
0.0000000000000000 0.0000000000000000 0.2707735090089329
0.5000000000000000 0.0000000000000000 0.2707735090089329
0.0000000000000000 0.5000000000000000 0.2707735090089329
0.5000000000000000 0.5000000000000000 0.2707735090089329
0.0000000000000000 0.7500000000000000 0.3127730853707575
0.7500000000000000 0.5000000000000000 0.3127730853707575
0.2500000000000000 0.0000000000000000 0.3127730853707575
0.5000000000000000 0.2500000000000000 0.3127730853707575
```

```
0.2500000000000000  0.5000000000000000  0.3127730853707575
0.0000000000000000  0.2500000000000000  0.3127730853707575
0.7500000000000000  0.0000000000000000  0.3127730853707575
0.5000000000000000  0.7500000000000000  0.3127730853707575
0.0000000000000000  0.0000000000000000  0.3547599910000017
0.0000000000000000  0.5000000000000000  0.3547599910000017
0.5000000000000000  0.0000000000000000  0.3547599910000017
0.5000000000000000  0.5000000000000000  0.3547599910000017
0.0000000000000000  0.7500000000000000  0.3966600050000011
0.2500000000000000  0.5000000000000000  0.3966600050000011
0.0000000000000000  0.2500000000000000  0.3966600050000011
0.7500000000000000  0.5000000000000000  0.3966600050000011
0.2500000000000000  0.0000000000000000  0.3966600050000011
0.7500000000000000  0.0000000000000000  0.3966600050000011
0.5000000000000000  0.2500000000000000  0.3966600050000011
0.5000000000000000  0.7500000000000000  0.3966600050000011
0.5000000000000000  0.0000000000000000  0.4385600199999971
0.0000000000000000  0.0000000000000000  0.4385600199999971
0.0000000000000000  0.5000000000000000  0.4385600199999971
0.5000000000000000  0.5000000000000000  0.4385600199999971
0.5000000000000000  0.2500000000000000  0.4806142007717991
0.2500000000000000  0.0000000000000000  0.4806142007717991
0.7500000000000000  0.5000000000000000  0.4806142007717991
0.0000000000000000  0.7500000000000000  0.4806142007717991
0.5000000000000000  0.7500000000000000  0.4806142007717991
0.7500000000000000  0.0000000000000000  0.4806142007717991
0.0000000000000000  0.2500000000000000  0.4806142007717991
0.2500000000000000  0.5000000000000000  0.4806142007717991
0.5000000000000000  0.0000000000000000  0.5226216488442800
0.0000000000000000  0.0000000000000000  0.5226216488442800
0.5000000000000000  0.5000000000000000  0.5226216488442800
0.0000000000000000  0.5000000000000000  0.5226216488442800
0.7500000000000000  0.0000000000000000  0.5646659412464932
0.2500000000000000  0.5000000000000000  0.5646659412464932
0.5000000000000000  0.2500000000000000  0.5646659412464932
0.2500000000000000  0.0000000000000000  0.5646659412464932
0.7500000000000000  0.5000000000000000  0.5646659412464932
0.0000000000000000  0.2500000000000000  0.5646659412464932
0.5000000000000000  0.7500000000000000  0.5646659412464932
0.0000000000000000  0.7500000000000000  0.5646659412464932
0.5000000000000000  0.5000000000000000  0.6066359701400675
0.0000000000000000  0.0000000000000000  0.6066359701400675
0.5000000000000000  0.0000000000000000  0.6066359701400675
0.0000000000000000  0.5000000000000000  0.6066359701400675
0.7500000000000000  0.5000000000000000  0.6488915096060550
0.5000000000000000  0.2500000000000000  0.6488915096060550
0.2500000000000000  0.5000000000000000  0.6488915096060550
0.0000000000000000  0.2500000000000000  0.6488915096060550
0.7500000000000000  0.0000000000000000  0.6488915096060550
0.5000000000000000  0.7500000000000000  0.6488915096060550
```

```
0.2500000000000000  0.0000000000000000  0.6488915096060550
0.0000000000000000  0.7500000000000000  0.6488915096060550
0.5000000000000000  0.5000000000000000  0.6898887332660033
0.0000000000000000  0.0000000000000000  0.6898887332660033
0.5000000000000000  0.0000000000000000  0.6898887332660033
0.0000000000000000  0.5000000000000000  0.6898887332660033
0.2500132527146198 -0.0000000000000000  0.7403711244411633
0.2499867472853802  0.5000000000000000  0.7403711244411633
0.7499867472853803 -0.0000000000000000  0.7403711244411633
0.7500132527146197  0.5000000000000000  0.7403711244411633
-0.0000000000000000  0.2499867472853802  0.7403711244411633
0.5000000000000000  0.2500132527146198  0.7403711244411633
-0.0000000000000000  0.7500132527146197  0.7403711244411633
0.5000000000000000  0.7499867472853803  0.7403711244411633
0.5000000000000000  0.5000000000000000  0.7397701965756600
0.0000000000000000  0.0000000000000000  0.7397701965756600
0.5000000000000000  0.0000000000000000  0.7397701965756600
0.0000000000000000  0.5000000000000000  0.7397701965756600
```

**KMgF₃ tetragonal**
  1.00000000000000
    7.9690999984999999    0.0000000000000000    0.0000000000000000
    0.0000000000000000    7.9690999984999999    0.0000000000000000
    0.0000000000000000    0.0000000000000000   47.8917007446000014
   K   Mg   F   Ag
   32   28   96   4
Direct
  0.2500000000000000  0.2500000000000000  0.1057676156251972
  0.2500000000000000  0.7499999699999975  0.1057676156251972
  0.7499999699999975  0.2500000000000000  0.1057676156251972
  0.7499999699999975  0.7499999699999975  0.1057676156251972
  0.2500000000000000  0.2500000000000000  0.1875557885213794
  0.2500000000000000  0.7499999699999975  0.1875557885213794
  0.7499999699999975  0.2500000000000000  0.1875557885213794
  0.7499999699999975  0.7499999699999975  0.1875557885213794
  0.2500000000000000  0.2500000000000000  0.2706867362157865
  0.2500000000000000  0.7499999699999975  0.2706867362157865
  0.7499999699999975  0.2500000000000000  0.2706867362157865
  0.7499999699999975  0.7499999699999975  0.2706867362157865
  0.2500000000000000  0.2500000000000000  0.3540000169999971
  0.2500000000000000  0.7499999699999975  0.3540000169999971
  0.7499999699999975  0.2500000000000000  0.3540000169999971
  0.7499999699999975  0.7499999699999975  0.3540000169999971
  0.2500000000000000  0.2500000000000000  0.4371999990000006
  0.2500000000000000  0.7499999699999975  0.4371999990000006
  0.7499999699999975  0.2500000000000000  0.4371999990000006
  0.7499999699999975  0.7499999699999975  0.4371999990000006
  0.2500000000000000  0.2500000000000000  0.5206393913345082
  0.2500000000000000  0.7499999699999975  0.5206393913345082
  0.7499999699999975  0.2500000000000000  0.5206393913345082
  0.7499999699999975  0.7499999699999975  0.5206393913345082
  0.2500000000000000  0.2500000000000000  0.6040855404009194
  0.2500000000000000  0.7499999699999975  0.6040855404009194
  0.7499999699999975  0.2500000000000000  0.6040855404009194
  0.7499999699999975  0.7499999699999975  0.6040855404009194
  0.2500000000000000  0.2500000000000000  0.6877962435853086
  0.2500000000000000  0.7499999699999975  0.6877962435853086
  0.7499999699999975  0.2500000000000000  0.6877962435853086
  0.7499999699999975  0.7499999699999975  0.6877962435853086
  0.0000000000000000  0.0000000000000000  0.1453458700474357
  0.0000000000000000  0.5000000000000000  0.1453458700474357
  0.5000000000000000  0.0000000000000000  0.1453458700474357
  0.5000000000000000  0.5000000000000000  0.1453458700474357
  0.0000000000000000  0.0000000000000000  0.2289564481792610
  0.0000000000000000  0.5000000000000000  0.2289564481792610
  0.5000000000000000  0.0000000000000000  0.2289564481792610
  0.5000000000000000  0.5000000000000000  0.2289564481792610
  0.0000000000000000  0.0000000000000000  0.3123225958626463
  0.0000000000000000  0.5000000000000000  0.3123225958626463

```
0.5000000000000000 0.0000000000000000 0.3123225958626463
0.5000000000000000 0.5000000000000000 0.3123225958626463
0.0000000000000000 0.0000000000000000 0.3956000080000024
0.0000000000000000 0.5000000000000000 0.3956000080000024
0.5000000000000000 0.0000000000000000 0.3956000080000024
0.5000000000000000 0.5000000000000000 0.3956000080000024
0.0000000000000000 0.0000000000000000 0.4789181215933953
0.0000000000000000 0.5000000000000000 0.4789181215933953
0.5000000000000000 0.0000000000000000 0.4789181215933953
0.5000000000000000 0.5000000000000000 0.4789181215933953
0.0000000000000000 0.0000000000000000 0.5623792884329080
0.0000000000000000 0.5000000000000000 0.5623792884329080
0.5000000000000000 0.0000000000000000 0.5623792884329080
0.5000000000000000 0.5000000000000000 0.5623792884329080
0.0000000000000000 0.0000000000000000 0.6458456321164425
0.0000000000000000 0.5000000000000000 0.6458456321164425
0.5000000000000000 0.0000000000000000 0.6458456321164425
0.5000000000000000 0.5000000000000000 0.6458456321164425
0.0000000000000000 0.5000000000000000 0.1040629470016796
0.0000000000000000 0.0000000000000000 0.1040629470016796
0.5000000000000000 0.0000000000000000 0.1040629470016796
0.5000000000000000 0.5000000000000000 0.1040629470016796
0.5000000000000000 0.2500000000000000 0.1458321217387635
0.7499999699999975 0.5000000000000000 0.1458321217387635
0.0000000000000000 0.2500000000000000 0.1458321217387635
0.2500000000000000 0.0000000000000000 0.1458321217387635
0.5000000000000000 0.7499999699999975 0.1458321217387635
0.2500000000000000 0.5000000000000000 0.1458321217387635
0.7499999699999975 0.0000000000000000 0.1458321217387635
0.0000000000000000 0.7499999699999975 0.1458321217387635
0.5000000000000000 0.0000000000000000 0.1872757298735763
0.5000000000000000 0.5000000000000000 0.1872757298735763
0.0000000000000000 0.0000000000000000 0.1872757298735763
0.0000000000000000 0.5000000000000000 0.1872757298735763
0.2500000000000000 0.5000000000000000 0.2290050752768877
0.7499999699999975 0.5000000000000000 0.2290050752768877
0.2500000000000000 0.0000000000000000 0.2290050752768877
0.0000000000000000 0.7499999699999975 0.2290050752768877
0.7499999699999975 0.0000000000000000 0.2290050752768877
0.5000000000000000 0.2500000000000000 0.2290050752768877
0.5000000000000000 0.7499999699999975 0.2290050752768877
0.0000000000000000 0.2500000000000000 0.2290050752768877
0.0000000000000000 0.0000000000000000 0.2706511912126928
0.5000000000000000 0.0000000000000000 0.2706511912126928
0.0000000000000000 0.5000000000000000 0.2706511912126928
0.5000000000000000 0.5000000000000000 0.2706511912126928
0.0000000000000000 0.7499999699999975 0.3123284649010714
0.7499999699999975 0.5000000000000000 0.3123284649010714
0.2500000000000000 0.0000000000000000 0.3123284649010714
0.5000000000000000 0.2500000000000000 0.3123284649010714
```

```
0.2500000000000000 0.5000000000000000 0.3123284649010714
0.0000000000000000 0.2500000000000000 0.3123284649010714
0.7499999699999975 0.0000000000000000 0.3123284649010714
0.5000000000000000 0.7499999699999975 0.3123284649010714
0.0000000000000000 0.0000000000000000 0.3540000169999971
0.0000000000000000 0.5000000000000000 0.3540000169999971
0.5000000000000000 0.0000000000000000 0.3540000169999971
0.5000000000000000 0.5000000000000000 0.3540000169999971
0.0000000000000000 0.7499999699999975 0.3956000080000024
0.2500000000000000 0.5000000000000000 0.3956000080000024
0.0000000000000000 0.2500000000000000 0.3956000080000024
0.7499999699999975 0.5000000000000000 0.3956000080000024
0.2500000000000000 0.0000000000000000 0.3956000080000024
0.7499999699999975 0.0000000000000000 0.3956000080000024
0.5000000000000000 0.2500000000000000 0.3956000080000024
0.5000000000000000 0.7499999699999975 0.3956000080000024
0.5000000000000000 0.0000000000000000 0.4371999990000006
0.0000000000000000 0.0000000000000000 0.4371999990000006
0.0000000000000000 0.5000000000000000 0.4371999990000006
0.5000000000000000 0.5000000000000000 0.4371999990000006
0.5000000000000000 0.2500000000000000 0.4789536319819437
0.2500000000000000 0.0000000000000000 0.4789536319819437
0.7499999699999975 0.5000000000000000 0.4789536319819437
0.0000000000000000 0.7499999699999975 0.4789536319819437
0.5000000000000000 0.7499999699999975 0.4789536319819437
0.7499999699999975 0.0000000000000000 0.4789536319819437
0.0000000000000000 0.2500000000000000 0.4789536319819437
0.2500000000000000 0.5000000000000000 0.4789536319819437
0.5000000000000000 0.0000000000000000 0.5206719198004657
0.0000000000000000 0.0000000000000000 0.5206719198004657
0.5000000000000000 0.5000000000000000 0.5206719198004657
0.0000000000000000 0.5000000000000000 0.5206719198004657
0.7499999699999975 0.0000000000000000 0.5624067369966386
0.2500000000000000 0.5000000000000000 0.5624067369966386
0.5000000000000000 0.2500000000000000 0.5624067369966386
0.2500000000000000 0.0000000000000000 0.5624067369966386
0.7499999699999975 0.5000000000000000 0.5624067369966386
0.0000000000000000 0.2500000000000000 0.5624067369966386
0.5000000000000000 0.7499999699999975 0.5624067369966386
0.0000000000000000 0.7499999699999975 0.5624067369966386
0.5000000000000000 0.5000000000000000 0.6041210953790989
0.0000000000000000 0.0000000000000000 0.6041210953790989
0.5000000000000000 0.0000000000000000 0.6041210953790989
0.0000000000000000 0.5000000000000000 0.6041210953790989
0.7499999699999975 0.5000000000000000 0.6458990658110692
0.5000000000000000 0.2500000000000000 0.6458990658110692
0.2500000000000000 0.5000000000000000 0.6458990658110692
0.0000000000000000 0.2500000000000000 0.6458990658110692
0.7499999699999975 0.0000000000000000 0.6458990658110692
0.5000000000000000 0.7499999699999975 0.6458990658110692
```

```
0.2500000000000000  0.0000000000000000  0.6458990658110692
0.0000000000000000  0.7499999699999975  0.6458990658110692
0.5000000000000000  0.5000000000000000  0.6872807780111440
0.0000000000000000  0.0000000000000000  0.6872807780111440
0.5000000000000000  0.0000000000000000  0.6872807780111440
0.0000000000000000  0.5000000000000000  0.6872807780111440
0.2500000000000000  0.0000000000000000  0.7371555463578332
0.2500000000000000  0.5000000000000000  0.7371555463578332
0.7500000000000000  0.0000000000000000  0.7371555463578332
0.7500000000000000  0.5000000000000000  0.7371555463578332
0.0000000000000000  0.2500000000000000  0.7371555463578332
0.5000000000000000  0.2500000000000000  0.7371555463578332
0.0000000000000000  0.7500000000000000  0.7371555463578332
0.5000000000000000  0.7500000000000000  0.7371555463578332
0.5000000000000000  0.5000000000000000  0.7375093028267062
0.0000000000000000  0.0000000000000000  0.7375093028267062
0.5000000000000000  0.0000000000000000  0.7375093028267062
0.0000000000000000  0.5000000000000000  0.7375093028267062
```

**KMgF₃ orthorhombic**
  1.00000000000000
    7.9690999984999999    0.0000000000000000    0.0000000000000000
    0.0000000000000000    7.9690999984999999    0.0000000000000000
    0.0000000000000000    0.0000000000000000   47.8917007446000014
   K  Mg  F  Ag
   32  28  96  4
Direct
  0.2500000000000000  0.2500000000000000  0.1057676156251972
  0.2500000000000000  0.7499999699999975  0.1057676156251972
  0.7499999699999975  0.2500000000000000  0.1057676156251972
  0.7499999699999975  0.7499999699999975  0.1057676156251972
  0.2500000000000000  0.2500000000000000  0.1875557885213794
  0.2500000000000000  0.7499999699999975  0.1875557885213794
  0.7499999699999975  0.2500000000000000  0.1875557885213794
  0.7499999699999975  0.7499999699999975  0.1875557885213794
  0.2500000000000000  0.2500000000000000  0.2706867362157865
  0.2500000000000000  0.7499999699999975  0.2706867362157865
  0.7499999699999975  0.2500000000000000  0.2706867362157865
  0.7499999699999975  0.7499999699999975  0.2706867362157865
  0.2500000000000000  0.2500000000000000  0.3540000169999971
  0.2500000000000000  0.7499999699999975  0.3540000169999971
  0.7499999699999975  0.2500000000000000  0.3540000169999971
  0.7499999699999975  0.7499999699999975  0.3540000169999971
  0.2500000000000000  0.2500000000000000  0.4371999990000006
  0.2500000000000000  0.7499999699999975  0.4371999990000006
  0.7499999699999975  0.2500000000000000  0.4371999990000006
  0.7499999699999975  0.7499999699999975  0.4371999990000006
  0.2500000000000000  0.2500000000000000  0.5206393913345082
  0.2500000000000000  0.7499999699999975  0.5206393913345082
  0.7499999699999975  0.2500000000000000  0.5206393913345082
  0.7499999699999975  0.7499999699999975  0.5206393913345082
  0.2500000000000000  0.2500000000000000  0.6040855404009194
  0.2500000000000000  0.7499999699999975  0.6040855404009194
  0.7499999699999975  0.2500000000000000  0.6040855404009194
  0.7499999699999975  0.7499999699999975  0.6040855404009194
  0.2500000000000000  0.2500000000000000  0.6877962435853086
  0.2500000000000000  0.7499999699999975  0.6877962435853086
  0.7499999699999975  0.2500000000000000  0.6877962435853086
  0.7499999699999975  0.7499999699999975  0.6877962435853086
  0.0000000000000000  0.0000000000000000  0.1453458700474357
  0.0000000000000000  0.5000000000000000  0.1453458700474357
  0.5000000000000000  0.0000000000000000  0.1453458700474357
  0.5000000000000000  0.5000000000000000  0.1453458700474357
  0.0000000000000000  0.0000000000000000  0.2289564481792610
  0.0000000000000000  0.5000000000000000  0.2289564481792610
  0.5000000000000000  0.0000000000000000  0.2289564481792610
  0.5000000000000000  0.5000000000000000  0.2289564481792610
  0.0000000000000000  0.0000000000000000  0.3123225958626463
  0.0000000000000000  0.5000000000000000  0.3123225958626463

```
0.5000000000000000  0.0000000000000000  0.3123225958626463
0.5000000000000000  0.5000000000000000  0.3123225958626463
0.0000000000000000  0.0000000000000000  0.3956000080000024
0.0000000000000000  0.5000000000000000  0.3956000080000024
0.5000000000000000  0.0000000000000000  0.3956000080000024
0.5000000000000000  0.5000000000000000  0.3956000080000024
0.0000000000000000  0.0000000000000000  0.4789181215933953
0.0000000000000000  0.5000000000000000  0.4789181215933953
0.5000000000000000  0.0000000000000000  0.4789181215933953
0.5000000000000000  0.5000000000000000  0.4789181215933953
0.0000000000000000  0.0000000000000000  0.5623792884329080
0.0000000000000000  0.5000000000000000  0.5623792884329080
0.5000000000000000  0.0000000000000000  0.5623792884329080
0.5000000000000000  0.5000000000000000  0.5623792884329080
0.0000000000000000  0.0000000000000000  0.6458456321164425
0.0000000000000000  0.5000000000000000  0.6458456321164425
0.5000000000000000  0.0000000000000000  0.6458456321164425
0.5000000000000000  0.5000000000000000  0.6458456321164425
0.0000000000000000  0.5000000000000000  0.1040629470016796
0.0000000000000000  0.0000000000000000  0.1040629470016796
0.5000000000000000  0.0000000000000000  0.1040629470016796
0.5000000000000000  0.5000000000000000  0.1040629470016796
0.5000000000000000  0.2500000000000000  0.1458321217387635
0.7499999699999975  0.5000000000000000  0.1458321217387635
0.0000000000000000  0.2500000000000000  0.1458321217387635
0.2500000000000000  0.0000000000000000  0.1458321217387635
0.5000000000000000  0.7499999699999975  0.1458321217387635
0.2500000000000000  0.5000000000000000  0.1458321217387635
0.7499999699999975  0.0000000000000000  0.1458321217387635
0.0000000000000000  0.7499999699999975  0.1458321217387635
0.5000000000000000  0.0000000000000000  0.1872757298735763
0.5000000000000000  0.5000000000000000  0.1872757298735763
0.0000000000000000  0.0000000000000000  0.1872757298735763
0.0000000000000000  0.5000000000000000  0.1872757298735763
0.2500000000000000  0.5000000000000000  0.2290050752768877
0.7499999699999975  0.5000000000000000  0.2290050752768877
0.2500000000000000  0.0000000000000000  0.2290050752768877
0.0000000000000000  0.7499999699999975  0.2290050752768877
0.7499999699999975  0.0000000000000000  0.2290050752768877
0.5000000000000000  0.2500000000000000  0.2290050752768877
0.5000000000000000  0.7499999699999975  0.2290050752768877
0.0000000000000000  0.2500000000000000  0.2290050752768877
0.0000000000000000  0.0000000000000000  0.2706511912126928
0.5000000000000000  0.0000000000000000  0.2706511912126928
0.0000000000000000  0.5000000000000000  0.2706511912126928
0.5000000000000000  0.5000000000000000  0.2706511912126928
0.0000000000000000  0.7499999699999975  0.3123284649010714
0.7499999699999975  0.5000000000000000  0.3123284649010714
0.2500000000000000  0.0000000000000000  0.3123284649010714
0.5000000000000000  0.2500000000000000  0.3123284649010714
```

```
0.2500000000000000 0.5000000000000000 0.3123284649010714
0.0000000000000000 0.2500000000000000 0.3123284649010714
0.7499999699999975 0.0000000000000000 0.3123284649010714
0.5000000000000000 0.7499999699999975 0.3123284649010714
0.0000000000000000 0.0000000000000000 0.3540000169999971
0.0000000000000000 0.5000000000000000 0.3540000169999971
0.5000000000000000 0.0000000000000000 0.3540000169999971
0.5000000000000000 0.5000000000000000 0.3540000169999971
0.0000000000000000 0.7499999699999975 0.3956000080000024
0.2500000000000000 0.5000000000000000 0.3956000080000024
0.0000000000000000 0.2500000000000000 0.3956000080000024
0.7499999699999975 0.5000000000000000 0.3956000080000024
0.2500000000000000 0.0000000000000000 0.3956000080000024
0.7499999699999975 0.0000000000000000 0.3956000080000024
0.5000000000000000 0.2500000000000000 0.3956000080000024
0.5000000000000000 0.7499999699999975 0.3956000080000024
0.5000000000000000 0.0000000000000000 0.4371999990000006
0.0000000000000000 0.0000000000000000 0.4371999990000006
0.0000000000000000 0.5000000000000000 0.4371999990000006
0.5000000000000000 0.5000000000000000 0.4371999990000006
0.5000000000000000 0.2500000000000000 0.4789536319819437
0.2500000000000000 0.0000000000000000 0.4789536319819437
0.7499999699999975 0.5000000000000000 0.4789536319819437
0.0000000000000000 0.7499999699999975 0.4789536319819437
0.5000000000000000 0.7499999699999975 0.4789536319819437
0.7499999699999975 0.0000000000000000 0.4789536319819437
0.0000000000000000 0.2500000000000000 0.4789536319819437
0.2500000000000000 0.5000000000000000 0.4789536319819437
0.5000000000000000 0.0000000000000000 0.5206719198004657
0.0000000000000000 0.0000000000000000 0.5206719198004657
0.5000000000000000 0.5000000000000000 0.5206719198004657
0.0000000000000000 0.5000000000000000 0.5206719198004657
0.7499999699999975 0.0000000000000000 0.5624067369966386
0.2500000000000000 0.5000000000000000 0.5624067369966386
0.5000000000000000 0.2500000000000000 0.5624067369966386
0.2500000000000000 0.0000000000000000 0.5624067369966386
0.7499999699999975 0.5000000000000000 0.5624067369966386
0.0000000000000000 0.2500000000000000 0.5624067369966386
0.5000000000000000 0.7499999699999975 0.5624067369966386
0.0000000000000000 0.7499999699999975 0.5624067369966386
0.5000000000000000 0.5000000000000000 0.6041210953790989
0.0000000000000000 0.0000000000000000 0.6041210953790989
0.5000000000000000 0.0000000000000000 0.6041210953790989
0.0000000000000000 0.5000000000000000 0.6041210953790989
0.7499999699999975 0.5000000000000000 0.6458990658110692
0.5000000000000000 0.2500000000000000 0.6458990658110692
0.2500000000000000 0.5000000000000000 0.6458990658110692
0.0000000000000000 0.2500000000000000 0.6458990658110692
0.7499999699999975 0.0000000000000000 0.6458990658110692
0.5000000000000000 0.7499999699999975 0.6458990658110692
```

```
0.2500000000000000  0.0000000000000000  0.6458990658110692
0.0000000000000000  0.7499999699999975  0.6458990658110692
0.5000000000000000  0.5000000000000000  0.6872807780111440
0.0000000000000000  0.0000000000000000  0.6872807780111440
0.5000000000000000  0.0000000000000000  0.6872807780111440
0.0000000000000000  0.5000000000000000  0.6872807780111440
0.2500000000000000  0.0000000000000000  0.7251555463578332
0.2500000000000000  0.5000000000000000  0.7551555463578332
0.7500000000000000  0.0000000000000000  0.7551555463578332
0.7500000000000000  0.5000000000000000  0.7251555463578332
0.0000000000000000  0.2500000000000000  0.7551555463578332
0.5000000000000000  0.2500000000000000  0.7251555463578332
0.0000000000000000  0.7500000000000000  0.7251555463578332
0.5000000000000000  0.7500000000000000  0.7551555463578332
0.5000000000000000  0.5000000000000000  0.7405093028267062
0.0000000000000000  0.0000000000000000  0.7405093028267062
0.5000000000000000  0.0000000000000000  0.7405093028267062
0.0000000000000000  0.5000000000000000  0.7405093028267062
```

**KCdF₃ tetragonal**
 1.00000000000000
     8.7931003571000002    0.0000000000000000    0.0000000000000000
     0.0000000000000000    8.7931003571000002    0.0000000000000000
     0.0000000000000000    0.0000000000000000   50.7757987975999967
   K   Cd   F   Ag
   32   28   96   4
Direct
  0.7500000000000000  0.7500000000000000  0.1057108455505613
  0.7500000000000000  0.2500000000000000  0.1057108455505613
  0.2500000000000000  0.7500000000000000  0.1057108455505613
  0.2500000000000000  0.2500000000000000  0.1057108455505613
  0.7500000000000000  0.7500000000000000  0.1858238940743533
  0.7500000000000000  0.2500000000000000  0.1858238940743533
  0.2500000000000000  0.7500000000000000  0.1858238940743533
  0.2500000000000000  0.2500000000000000  0.1858238940743533
  0.7500000000000000  0.7500000000000000  0.2714811693575498
  0.7500000000000000  0.2500000000000000  0.2714811693575498
  0.2500000000000000  0.7500000000000000  0.2714811693575498
  0.2500000000000000  0.2500000000000000  0.2714811693575498
  0.7500000000000000  0.7500000000000000  0.3582300110000034
  0.7500000000000000  0.2500000000000000  0.3582300110000034
  0.2500000000000000  0.7500000000000000  0.3582300110000034
  0.2500000000000000  0.2500000000000000  0.3582300110000034
  0.7500000000000000  0.7500000000000000  0.4448199859999988
  0.7500000000000000  0.2500000000000000  0.4448199859999988
  0.2500000000000000  0.7500000000000000  0.4448199859999988
  0.2500000000000000  0.2500000000000000  0.4448199859999988
  0.7500000000000000  0.7500000000000000  0.5341004491896316
  0.7500000000000000  0.2500000000000000  0.5341004491896316
  0.2500000000000000  0.7500000000000000  0.5341004491896316
  0.2500000000000000  0.2500000000000000  0.5341004491896316
  0.7500000000000000  0.7500000000000000  0.6225520244263012
  0.7500000000000000  0.2500000000000000  0.6225520244263012
  0.2500000000000000  0.7500000000000000  0.6225520244263012
  0.2500000000000000  0.2500000000000000  0.6225520244263012
  0.7500000000000000  0.7500000000000000  0.7121517431295437
  0.7500000000000000  0.2500000000000000  0.7121517431295437
  0.2500000000000000  0.7500000000000000  0.7121517431295437
  0.2500000000000000  0.2500000000000000  0.7121517431295437
  0.0000000000000000  0.0000000000000000  0.1408346093264967
  0.0000000000000000  0.5000000000000000  0.1408346093264967
  0.5000000000000000  0.0000000000000000  0.1408346093264967
  0.5000000000000000  0.5000000000000000  0.1408346093264967
  0.0000000000000000  0.0000000000000000  0.2278916587290070
  0.0000000000000000  0.5000000000000000  0.2278916587290070
  0.5000000000000000  0.0000000000000000  0.2278916587290070
  0.5000000000000000  0.5000000000000000  0.2278916587290070
  0.0000000000000000  0.0000000000000000  0.3147679302584265
  0.0000000000000000  0.5000000000000000  0.3147679302584265

```
0.5000000000000000 0.0000000000000000 0.3147679302584265
0.5000000000000000 0.5000000000000000 0.3147679302584265
0.0000000000000000 0.0000000000000000 0.4015299940000006
0.0000000000000000 0.5000000000000000 0.4015299940000006
0.5000000000000000 0.0000000000000000 0.4015299940000006
0.5000000000000000 0.5000000000000000 0.4015299940000006
0.0000000000000000 0.0000000000000000 0.4882991693518156
0.0000000000000000 0.5000000000000000 0.4882991693518156
0.5000000000000000 0.0000000000000000 0.4882991693518156
0.5000000000000000 0.5000000000000000 0.4882991693518156
0.0000000000000000 0.0000000000000000 0.5751556587767703
0.0000000000000000 0.5000000000000000 0.5751556587767703
0.5000000000000000 0.0000000000000000 0.5751556587767703
0.5000000000000000 0.5000000000000000 0.5751556587767703
0.0000000000000000 0.0000000000000000 0.6621099880319838
0.0000000000000000 0.5000000000000000 0.6621099880319838
0.5000000000000000 0.0000000000000000 0.6621099880319838
0.5000000000000000 0.5000000000000000 0.6621099880319838
0.0000000000000000 0.0000000000000000 0.0981816797369746
0.5000000000000000 0.5000000000000000 0.0981816797369746
0.5000000000000000 0.0000000000000000 0.0981816797369746
0.0000000000000000 0.5000000000000000 0.0981816797369746
0.5000000000000000 0.7500000000000000 0.1424678727494921
0.7500000000000000 0.5000000000000000 0.1424678727494921
0.2500000000000000 0.0000000000000000 0.1424678727494921
0.0000000000000000 0.2500000000000000 0.1424678727494921
0.2500000000000000 0.5000000000000000 0.1424678727494921
0.5000000000000000 0.2500000000000000 0.1424678727494921
0.0000000000000000 0.7500000000000000 0.1424678727494921
0.7500000000000000 0.0000000000000000 0.1424678727494921
0.5000000000000000 0.0000000000000000 0.1845679707296384
0.5000000000000000 0.5000000000000000 0.1845679707296384
0.0000000000000000 0.5000000000000000 0.1845679707296384
0.0000000000000000 0.0000000000000000 0.1845679707296384
0.2500000000000000 0.5000000000000000 0.2280820128390246
0.7500000000000000 0.5000000000000000 0.2280820128390246
0.0000000000000000 0.7500000000000000 0.2280820128390246
0.5000000000000000 0.2500000000000000 0.2280820128390246
0.7500000000000000 0.0000000000000000 0.2280820128390246
0.5000000000000000 0.7500000000000000 0.2280820128390246
0.0000000000000000 0.2500000000000000 0.2280820128390246
0.2500000000000000 0.0000000000000000 0.2280820128390246
0.0000000000000000 0.5000000000000000 0.2713486312363855
0.5000000000000000 0.0000000000000000 0.2713486312363855
0.5000000000000000 0.5000000000000000 0.2713486312363855
0.0000000000000000 0.0000000000000000 0.2713486312363855
0.0000000000000000 0.2500000000000000 0.3147565938856545
0.7500000000000000 0.0000000000000000 0.3147565938856545
0.2500000000000000 0.0000000000000000 0.3147565938856545
0.7500000000000000 0.5000000000000000 0.3147565938856545
```

```
0.5000000000000000 0.7500000000000000 0.3147565938856545
0.2500000000000000 0.5000000000000000 0.3147565938856545
0.5000000000000000 0.2500000000000000 0.3147565938856545
0.0000000000000000 0.7500000000000000 0.3147565938856545
0.0000000000000000 0.0000000000000000 0.3582300110000034
0.0000000000000000 0.5000000000000000 0.3582300110000034
0.5000000000000000 0.5000000000000000 0.3582300110000034
0.5000000000000000 0.0000000000000000 0.3582300110000034
0.0000000000000000 0.7500000000000000 0.4015299940000006
0.7500000000000000 0.5000000000000000 0.4015299940000006
0.7500000000000000 0.0000000000000000 0.4015299940000006
0.2500000000000000 0.5000000000000000 0.4015299940000006
0.2500000000000000 0.0000000000000000 0.4015299940000006
0.0000000000000000 0.2500000000000000 0.4015299940000006
0.5000000000000000 0.7500000000000000 0.4015299940000006
0.5000000000000000 0.2500000000000000 0.4015299940000006
0.0000000000000000 0.0000000000000000 0.4448199859999988
0.0000000000000000 0.5000000000000000 0.4448199859999988
0.5000000000000000 0.5000000000000000 0.4448199859999988
0.5000000000000000 0.0000000000000000 0.4448199859999988
0.2500000000000000 0.0000000000000000 0.4872352209567390
0.2500000000000000 0.5000000000000000 0.4872352209567390
0.0000000000000000 0.7500000000000000 0.4872352209567390
0.0000000000000000 0.2500000000000000 0.4872352209567390
0.7500000000000000 0.0000000000000000 0.4872352209567390
0.5000000000000000 0.2500000000000000 0.4872352209567390
0.7500000000000000 0.5000000000000000 0.4872352209567390
0.5000000000000000 0.7500000000000000 0.4872352209567390
0.0000000000000000 0.5000000000000000 0.5315225602339980
0.0000000000000000 0.0000000000000000 0.5315225602339980
0.5000000000000000 0.0000000000000000 0.5315225602339980
0.5000000000000000 0.5000000000000000 0.5315225602339980
0.7500000000000000 0.0000000000000000 0.5731637390193752
0.0000000000000000 0.7500000000000000 0.5731637390193752
0.7500000000000000 0.5000000000000000 0.5731637390193752
0.0000000000000000 0.2500000000000000 0.5731637390193752
0.2500000000000000 0.0000000000000000 0.5731637390193752
0.5000000000000000 0.7500000000000000 0.5731637390193752
0.2500000000000000 0.5000000000000000 0.5731637390193752
0.5000000000000000 0.2500000000000000 0.5731637390193752
0.5000000000000000 0.0000000000000000 0.6183233951059000
0.0000000000000000 0.0000000000000000 0.6183233951059000
0.5000000000000000 0.5000000000000000 0.6183233951059000
0.0000000000000000 0.5000000000000000 0.6183233951059000
0.0000000000000000 0.7500000000000000 0.6593454470339122
0.0000000000000000 0.2500000000000000 0.6593454470339122
0.2500000000000000 0.0000000000000000 0.6593454470339122
0.7500000000000000 0.0000000000000000 0.6593454470339122
0.5000000000000000 0.7500000000000000 0.6593454470339122
0.7500000000000000 0.5000000000000000 0.6593454470339122
```

```
0.2500000000000000  0.5000000000000000  0.6593454470339122
0.5000000000000000  0.2500000000000000  0.6593454470339122
0.0000000000000000  0.5000000000000000  0.7049531128816616
0.0000000000000000  0.0000000000000000  0.7049531128816616
0.5000000000000000  0.5000000000000000  0.7049531128816616
0.5000000000000000  0.0000000000000000  0.7049531128816616
0.2500000000000000  0.0000000000000000  0.7480056146962872
0.2500000000000000  0.5000000000000000  0.7480056146962872
0.7500000000000000  0.0000000000000000  0.7480056146962872
0.7500000000000000  0.5000000000000000  0.7480056146962872
0.0000000000000000  0.2500000000000000  0.7480056146962872
0.5000000000000000  0.2500000000000000  0.7480056146962872
0.0000000000000000  0.7500000000000000  0.7480056146962872
0.5000000000000000  0.7500000000000000  0.7480056146962872
0.5000000000000000  0.5000000000000000  0.7495531945390612
0.0000000000000000  0.0000000000000000  0.7495531945390612
0.5000000000000000  0.0000000000000000  0.7495531945390612
0.0000000000000000  0.5000000000000000  0.7495531945390612
```

**KCdF₃ orthorhombic**
  1.00000000000000
    8.7931003571000002   0.0000000000000000   0.0000000000000000
    0.0000000000000000   8.7931003571000002   0.0000000000000000
    0.0000000000000000   0.0000000000000000  50.7757987975999967
   K   Cd   F   Ag
   32   28   96   4
Direct
 0.7500000000000000  0.7500000000000000  0.1057108455505613
 0.7500000000000000  0.2500000000000000  0.1057108455505613
 0.2500000000000000  0.7500000000000000  0.1057108455505613
 0.2500000000000000  0.2500000000000000  0.1057108455505613
 0.7500000000000000  0.7500000000000000  0.1858238940743533
 0.7500000000000000  0.2500000000000000  0.1858238940743533
 0.2500000000000000  0.7500000000000000  0.1858238940743533
 0.2500000000000000  0.2500000000000000  0.1858238940743533
 0.7500000000000000  0.7500000000000000  0.2714811693575498
 0.7500000000000000  0.2500000000000000  0.2714811693575498
 0.2500000000000000  0.7500000000000000  0.2714811693575498
 0.2500000000000000  0.2500000000000000  0.2714811693575498
 0.7500000000000000  0.7500000000000000  0.3582300110000034
 0.7500000000000000  0.2500000000000000  0.3582300110000034
 0.2500000000000000  0.7500000000000000  0.3582300110000034
 0.2500000000000000  0.2500000000000000  0.3582300110000034
 0.7500000000000000  0.7500000000000000  0.4448199859999988
 0.7500000000000000  0.2500000000000000  0.4448199859999988
 0.2500000000000000  0.7500000000000000  0.4448199859999988
 0.2500000000000000  0.2500000000000000  0.4448199859999988
 0.7500000000000000  0.7500000000000000  0.5340567555380602
 0.7500000000000000  0.2500000000000000  0.5340567555380602
 0.2500000000000000  0.7500000000000000  0.5340567555380602
 0.2500000000000000  0.2500000000000000  0.5340567555380602
 0.7500000000000000  0.7500000000000000  0.6224425181079276
 0.7500000000000000  0.2500000000000000  0.6224425181079276
 0.2500000000000000  0.7500000000000000  0.6224425181079276
 0.2500000000000000  0.2500000000000000  0.6224425181079276
 0.7500000000000000  0.7500000000000000  0.7119436248095955
 0.7500000000000000  0.2500000000000000  0.7119436248095955
 0.2500000000000000  0.7500000000000000  0.7119436248095955
 0.2500000000000000  0.2500000000000000  0.7119436248095955
 0.0000000000000000  0.0000000000000000  0.1408346093264967
 0.0000000000000000  0.5000000000000000  0.1408346093264967
 0.5000000000000000  0.0000000000000000  0.1408346093264967
 0.5000000000000000  0.5000000000000000  0.1408346093264967
 0.0000000000000000  0.0000000000000000  0.2278916587290070
 0.0000000000000000  0.5000000000000000  0.2278916587290070
 0.5000000000000000  0.0000000000000000  0.2278916587290070
 0.5000000000000000  0.5000000000000000  0.2278916587290070
 0.0000000000000000  0.0000000000000000  0.3147679302584265
 0.0000000000000000  0.5000000000000000  0.3147679302584265

```
0.5000000000000000 0.0000000000000000 0.3147679302584265
0.5000000000000000 0.5000000000000000 0.3147679302584265
0.0000000000000000 0.0000000000000000 0.4015299940000006
0.0000000000000000 0.5000000000000000 0.4015299940000006
0.5000000000000000 0.0000000000000000 0.4015299940000006
0.5000000000000000 0.5000000000000000 0.4015299940000006
0.0000000000000000 0.0000000000000000 0.4882861592603348
0.0000000000000000 0.5000000000000000 0.4882861592603348
0.5000000000000000 0.0000000000000000 0.4882861592603348
0.5000000000000000 0.5000000000000000 0.4882861592603348
0.0000000000000000 0.0000000000000000 0.5751008099874808
0.0000000000000000 0.5000000000000000 0.5751008099874808
0.5000000000000000 0.0000000000000000 0.5751008099874808
0.5000000000000000 0.5000000000000000 0.5751008099874808
0.0000000000000000 0.0000000000000000 0.6620109355571986
0.0000000000000000 0.5000000000000000 0.6620109355571986
0.5000000000000000 0.0000000000000000 0.6620109355571986
0.5000000000000000 0.5000000000000000 0.6620109355571986
0.0000000000000000 0.0000000000000000 0.0981816797369746
0.5000000000000000 0.5000000000000000 0.0981816797369746
0.5000000000000000 0.0000000000000000 0.0981816797369746
0.0000000000000000 0.5000000000000000 0.0981816797369746
0.5000000000000000 0.7500000000000000 0.1424678727494921
0.7500000000000000 0.5000000000000000 0.1424678727494921
0.2500000000000000 0.0000000000000000 0.1424678727494921
0.0000000000000000 0.2500000000000000 0.1424678727494921
0.2500000000000000 0.5000000000000000 0.1424678727494921
0.5000000000000000 0.2500000000000000 0.1424678727494921
0.0000000000000000 0.7500000000000000 0.1424678727494921
0.7500000000000000 0.0000000000000000 0.1424678727494921
0.5000000000000000 0.0000000000000000 0.1845679707296384
0.5000000000000000 0.5000000000000000 0.1845679707296384
0.0000000000000000 0.5000000000000000 0.1845679707296384
0.0000000000000000 0.0000000000000000 0.1845679707296384
0.2500000000000000 0.5000000000000000 0.2280820128390246
0.7500000000000000 0.5000000000000000 0.2280820128390246
0.0000000000000000 0.7500000000000000 0.2280820128390246
0.5000000000000000 0.2500000000000000 0.2280820128390246
0.7500000000000000 0.0000000000000000 0.2280820128390246
0.5000000000000000 0.7500000000000000 0.2280820128390246
0.0000000000000000 0.2500000000000000 0.2280820128390246
0.2500000000000000 0.0000000000000000 0.2280820128390246
0.0000000000000000 0.5000000000000000 0.2713486312363855
0.5000000000000000 0.0000000000000000 0.2713486312363855
0.5000000000000000 0.5000000000000000 0.2713486312363855
0.0000000000000000 0.0000000000000000 0.2713486312363855
0.0000000000000000 0.2500000000000000 0.3147565938856545
0.7500000000000000 0.0000000000000000 0.3147565938856545
0.2500000000000000 0.0000000000000000 0.3147565938856545
0.7500000000000000 0.5000000000000000 0.3147565938856545
```

```
0.5000000000000000 0.7500000000000000 0.3147565938856545
0.2500000000000000 0.5000000000000000 0.3147565938856545
0.5000000000000000 0.2500000000000000 0.3147565938856545
0.0000000000000000 0.7500000000000000 0.3147565938856545
0.0000000000000000 0.0000000000000000 0.3582300110000034
0.0000000000000000 0.5000000000000000 0.3582300110000034
0.5000000000000000 0.5000000000000000 0.3582300110000034
0.5000000000000000 0.0000000000000000 0.3582300110000034
0.0000000000000000 0.7500000000000000 0.4015299940000006
0.7500000000000000 0.5000000000000000 0.4015299940000006
0.7500000000000000 0.0000000000000000 0.4015299940000006
0.2500000000000000 0.5000000000000000 0.4015299940000006
0.2500000000000000 0.0000000000000000 0.4015299940000006
0.0000000000000000 0.2500000000000000 0.4015299940000006
0.5000000000000000 0.7500000000000000 0.4015299940000006
0.5000000000000000 0.2500000000000000 0.4015299940000006
0.0000000000000000 0.0000000000000000 0.4448199859999988
0.0000000000000000 0.5000000000000000 0.4448199859999988
0.5000000000000000 0.5000000000000000 0.4448199859999988
0.5000000000000000 0.0000000000000000 0.4448199859999988
0.2500000000000000 0.0000000000000000 0.4872370160817766
0.2500000000000000 0.5000000000000000 0.4872390206738765
0.0000000000000000 0.7500000000000000 0.4872370160817766
0.0000000000000000 0.2500000000000000 0.4872390206738765
0.7500000000000000 0.0000000000000000 0.4872390206738765
0.5000000000000000 0.2500000000000000 0.4872370160817766
0.7500000000000000 0.5000000000000000 0.4872370160817766
0.5000000000000000 0.7500000000000000 0.4872390206738765
0.0000000000000000 0.5000000000000000 0.5314897342500737
0.0000000000000000 0.0000000000000000 0.5314897342500737
0.5000000000000000 0.0000000000000000 0.5314897342500737
0.5000000000000000 0.5000000000000000 0.5314897342500737
0.7500000000000000 0.0000000000000000 0.5731289566839139
0.0000000000000000 0.7500000000000000 0.5731347743691796
0.7500000000000000 0.5000000000000000 0.5731347743691796
0.0000000000000000 0.2500000000000000 0.5731289566839139
0.2500000000000000 0.0000000000000000 0.5731347743691796
0.5000000000000000 0.7500000000000000 0.5731289566839139
0.2500000000000000 0.5000000000000000 0.5731289566839139
0.5000000000000000 0.2500000000000000 0.5731347743691796
0.5000000000000000 0.0000000000000000 0.6182568037938658
0.0000000000000000 0.0000000000000000 0.6182568037938658
0.5000000000000000 0.5000000000000000 0.6182568037938658
0.0000000000000000 0.5000000000000000 0.6182568037938658
0.0000000000000000 0.7500000000000000 0.6592551776519392
0.0000000000000000 0.2500000000000000 0.6592538101558210
0.2500000000000000 0.0000000000000000 0.6592551776519392
0.7500000000000000 0.0000000000000000 0.6592538101558210
0.5000000000000000 0.7500000000000000 0.6592538101558210
0.7500000000000000 0.5000000000000000 0.6592551776519392
```

```
0.2500000000000000  0.5000000000000000  0.6592538101558210
0.5000000000000000  0.2500000000000000  0.6592551776519392
0.0000000000000000  0.5000000000000000  0.7048362725659822
0.0000000000000000  0.0000000000000000  0.7048362725659822
0.5000000000000000  0.5000000000000000  0.7048362725659822
0.5000000000000000  0.0000000000000000  0.7048362725659822
0.2500000000000000  0.0000000000000000  0.7478431162921476
0.2500000000000000  0.5000000000000000  0.7478466863885198
0.7500000000000000  0.0000000000000000  0.7478466863885198
0.7500000000000000  0.5000000000000000  0.7478431162921476
0.0000000000000000  0.2500000000000000  0.7478466863885198
0.5000000000000000  0.2500000000000000  0.7478431162921476
0.0000000000000000  0.7500000000000000  0.7478431162921476
0.5000000000000000  0.7500000000000000  0.7478466863885198
0.5000000000000000  0.5000000000000000  0.7494204409381131
0.0000000000000000  0.0000000000000000  0.7494204409381131
0.5000000000000000  0.0000000000000000  0.7494204409381131
0.0000000000000000  0.5000000000000000  0.7494204409381131
```

**KCdF₃ quasi-molecular**
  1.00000000000000
    8.7931003571000002    0.0000000000000000    0.0000000000000000
    0.0000000000000000    8.7931003571000002    0.0000000000000000
    0.0000000000000000    0.0000000000000000   50.7757987975999967
   K   Cd   F   Ag
   32   28   96   4
Direct
  0.7500000000000000  0.7500000000000000  0.1057108455505613
  0.7500000000000000  0.2500000000000000  0.1057108455505613
  0.2500000000000000  0.7500000000000000  0.1057108455505613
  0.2500000000000000  0.2500000000000000  0.1057108455505613
  0.7500000000000000  0.7500000000000000  0.1858238940743533
  0.7500000000000000  0.2500000000000000  0.1858238940743533
  0.2500000000000000  0.7500000000000000  0.1858238940743533
  0.2500000000000000  0.2500000000000000  0.1858238940743533
  0.7500000000000000  0.7500000000000000  0.2714811693575498
  0.7500000000000000  0.2500000000000000  0.2714811693575498
  0.2500000000000000  0.7500000000000000  0.2714811693575498
  0.2500000000000000  0.2500000000000000  0.2714811693575498
  0.7500000000000000  0.7500000000000000  0.3582300110000034
  0.7500000000000000  0.2500000000000000  0.3582300110000034
  0.2500000000000000  0.7500000000000000  0.3582300110000034
  0.2500000000000000  0.2500000000000000  0.3582300110000034
  0.7500000000000000  0.7500000000000000  0.4448199859999988
  0.7500000000000000  0.2500000000000000  0.4448199859999988
  0.2500000000000000  0.7500000000000000  0.4448199859999988
  0.2500000000000000  0.2500000000000000  0.4448199859999988
  0.7500000000000000  0.7500000000000000  0.5316817031871532
  0.7500000000000000  0.2500000000000000  0.5316817031871532
  0.2500000000000000  0.7500000000000000  0.5316817031871532
  0.2500000000000000  0.2500000000000000  0.5316817031871532
  0.7500000000000000  0.7500000000000000  0.6188925922766626
  0.7500000000000000  0.2500000000000000  0.6188925922766626
  0.2500000000000000  0.7500000000000000  0.6188925922766626
  0.2500000000000000  0.2500000000000000  0.6188925922766626
  0.7500000000000000  0.7500000000000000  0.7080798518123165
  0.7500000000000000  0.2500000000000000  0.7080798518123165
  0.2500000000000000  0.7500000000000000  0.7080798518123165
  0.2500000000000000  0.2500000000000000  0.7080798518123165
  0.0000000000000000  0.0000000000000000  0.1408346093264967
  0.0000000000000000  0.5000000000000000  0.1408346093264967
  0.5000000000000000  0.0000000000000000  0.1408346093264967
  0.5000000000000000  0.5000000000000000  0.1408346093264967
  0.0000000000000000  0.0000000000000000  0.2278916587290070
  0.0000000000000000  0.5000000000000000  0.2278916587290070
  0.5000000000000000  0.0000000000000000  0.2278916587290070
  0.5000000000000000  0.5000000000000000  0.2278916587290070
  0.0000000000000000  0.0000000000000000  0.3147679302584265
  0.0000000000000000  0.5000000000000000  0.3147679302584265

```
0.5000000000000000 0.0000000000000000 0.3147679302584265
0.5000000000000000 0.5000000000000000 0.3147679302584265
0.0000000000000000 0.0000000000000000 0.4015299940000006
0.0000000000000000 0.5000000000000000 0.4015299940000006
0.5000000000000000 0.0000000000000000 0.4015299940000006
0.5000000000000000 0.5000000000000000 0.4015299940000006
0.0000000000000000 0.0000000000000000 0.4882190539374934
0.0000000000000000 0.5000000000000000 0.4882190539374934
0.5000000000000000 0.0000000000000000 0.4882190539374934
0.5000000000000000 0.5000000000000000 0.4882190539374934
0.0000000000000000 0.0000000000000000 0.5750126976314435
0.0000000000000000 0.5000000000000000 0.5750126976314434
0.5000000000000000 0.0000000000000000 0.5750126976314435
0.5000000000000000 0.5000000000000000 0.5750126976314435
0.0000000000000000 0.0000000000000000 0.6617132914090084
0.0000000000000000 0.5000000000000000 0.6617132914090084
0.5000000000000000 0.0000000000000000 0.6617132914090084
0.5000000000000000 0.5000000000000000 0.6617132914090084
0.0000000000000000 0.0000000000000000 0.0981816797369746
0.5000000000000000 0.5000000000000000 0.0981816797369746
0.5000000000000000 0.0000000000000000 0.0981816797369746
0.0000000000000000 0.5000000000000000 0.0981816797369746
0.5000000000000000 0.7500000000000000 0.1424678727494921
0.7500000000000000 0.5000000000000000 0.1424678727494921
0.2500000000000000 0.0000000000000000 0.1424678727494921
0.0000000000000000 0.2500000000000000 0.1424678727494921
0.2500000000000000 0.5000000000000000 0.1424678727494921
0.5000000000000000 0.2500000000000000 0.1424678727494921
0.0000000000000000 0.7500000000000000 0.1424678727494921
0.7500000000000000 0.0000000000000000 0.1424678727494921
0.5000000000000000 0.0000000000000000 0.1845679707296384
0.5000000000000000 0.5000000000000000 0.1845679707296384
0.0000000000000000 0.5000000000000000 0.1845679707296384
0.0000000000000000 0.0000000000000000 0.1845679707296384
0.2500000000000000 0.5000000000000000 0.2280820128390246
0.7500000000000000 0.5000000000000000 0.2280820128390246
0.0000000000000000 0.7500000000000000 0.2280820128390246
0.5000000000000000 0.2500000000000000 0.2280820128390246
0.7500000000000000 0.0000000000000000 0.2280820128390246
0.5000000000000000 0.7500000000000000 0.2280820128390246
0.0000000000000000 0.2500000000000000 0.2280820128390246
0.2500000000000000 0.0000000000000000 0.2280820128390246
0.0000000000000000 0.5000000000000000 0.2713486312363855
0.5000000000000000 0.0000000000000000 0.2713486312363855
0.5000000000000000 0.5000000000000000 0.2713486312363855
0.0000000000000000 0.0000000000000000 0.2713486312363855
0.0000000000000000 0.2500000000000000 0.3147565938856545
0.7500000000000000 0.0000000000000000 0.3147565938856545
0.2500000000000000 0.0000000000000000 0.3147565938856545
0.7500000000000000 0.5000000000000000 0.3147565938856545
```

```
0.5000000000000000 0.7500000000000000 0.3147565938856545
0.2500000000000000 0.5000000000000000 0.3147565938856545
0.5000000000000000 0.2500000000000000 0.3147565938856545
0.0000000000000000 0.7500000000000000 0.3147565938856545
0.0000000000000000 0.0000000000000000 0.3582300110000034
0.0000000000000000 0.5000000000000000 0.3582300110000034
0.5000000000000000 0.5000000000000000 0.3582300110000034
0.5000000000000000 0.0000000000000000 0.3582300110000034
0.0000000000000000 0.7500000000000000 0.4015299940000006
0.7500000000000000 0.5000000000000000 0.4015299940000006
0.7500000000000000 0.0000000000000000 0.4015299940000006
0.2500000000000000 0.5000000000000000 0.4015299940000006
0.2500000000000000 0.0000000000000000 0.4015299940000006
0.0000000000000000 0.2500000000000000 0.4015299940000006
0.5000000000000000 0.7500000000000000 0.4015299940000006
0.5000000000000000 0.2500000000000000 0.4015299940000006
0.0000000000000000 0.0000000000000000 0.4448199859999988
0.0000000000000000 0.5000000000000000 0.4448199859999988
0.5000000000000000 0.5000000000000000 0.4448199859999988
0.5000000000000000 0.0000000000000000 0.4448199859999988
0.2500000000000000 0.0000000000000000 0.4881952730200473
0.2500000000000000 0.5000000000000000 0.4881952730200473
0.0000000000000000 0.7500000000000000 0.4881952730200473
0.0000000000000000 0.2500000000000000 0.4881952730200473
0.7500000000000000 0.0000000000000000 0.4881952730200473
0.5000000000000000 0.2500000000000000 0.4881952730200473
0.7500000000000000 0.5000000000000000 0.4881952730200473
0.5000000000000000 0.7500000000000000 0.4881952730200473
0.0000000000000000 0.5000000000000000 0.5316229560802577
0.0000000000000000 0.0000000000000000 0.5316229560802577
0.5000000000000000 0.0000000000000000 0.5316229560802577
0.5000000000000000 0.5000000000000000 0.5316229560802577
0.7500000000000000 0.0000000000000000 0.5749799567906198
0.0000000000000000 0.7500000000000000 0.5749799567906198
0.7500000000000000 0.5000000000000000 0.5749799567906198
0.0000000000000000 0.2500000000000000 0.5749799567906198
0.2500000000000000 0.0000000000000000 0.5749799567906198
0.5000000000000000 0.7500000000000000 0.5749799567906198
0.2500000000000000 0.5000000000000000 0.5749799567906198
0.5000000000000000 0.2500000000000000 0.5749799567906198
0.5000000000000000 0.0000000000000000 0.6184610509193981
0.0000000000000000 0.0000000000000000 0.6184610509193981
0.5000000000000000 0.5000000000000000 0.6184610509193981
0.0000000000000000 0.5000000000000000 0.6184610509193981
0.0000000000000000 0.7500000000000000 0.6614592233746542
0.0000000000000000 0.2500000000000000 0.6614592233746542
0.2500000000000000 0.0000000000000000 0.6614592233746542
0.7500000000000000 0.0000000000000000 0.6614592233746542
0.5000000000000000 0.7500000000000000 0.6614592233746542
0.7500000000000000 0.5000000000000000 0.6614592233746542
```

```
0.2500000000000000  0.5000000000000000  0.6614592233746542
0.5000000000000000  0.2500000000000000  0.6614592233746542
0.0000000000000000  0.5000000000000000  0.7049658866304781
0.0000000000000000  0.0000000000000000  0.7049658866304781
0.5000000000000000  0.5000000000000000  0.7049658866304781
0.5000000000000000  0.0000000000000000  0.7049658866304781
0.2687704261608667  0.0000000000000000  0.7465384941986893
0.2312295738391335  0.5000000000000000  0.7465384941986893
0.7312295738391338  0.0000000000000000  0.7465384941986893
0.7687704261608662  0.5000000000000000  0.7465384941986893
0.0000000000000000  0.2312295738391335  0.7465384941986893
0.5000000000000000  0.2687704261608667  0.7465384941986893
0.0000000000000000  0.7687704261608662  0.7465384941986893
0.5000000000000000  0.7312295738391338  0.7465384941986893
0.5000000000000000  0.5000000000000000  0.7475747569258511
0.0000000000000000  0.0000000000000000  0.7475747569258511
0.5000000000000000  0.0000000000000000  0.7475747569258511
0.0000000000000000  0.5000000000000000  0.7475747569258511
```

**MgO tetragonal**
   1.00000000000000
     8.4191999435000007    0.0000000000000000    0.0000000000000000
     0.0000000000000000    8.4191999435000007    0.0000000000000000
     0.0000000000000000    0.0000000000000000   41.0480995177999972
   Mg   O   F   Ag
    88   88   8   4
Direct
  0.2500000000000000  0.0000000000000000  0.1222401265431685
  0.5000000000000000  0.7499999719999977  0.1222401265431685
  0.0000000000000000  0.2500000000000000  0.1222401265431685
  0.7499999719999977  0.5000000000000000  0.1222401265431685
  0.2500000000000000  0.5000000000000000  0.1222401265431685
  0.5000000000000000  0.2500000000000000  0.1222401265431685
  0.0000000000000000  0.7499999719999977  0.1222401265431685
  0.7499999719999977  0.0000000000000000  0.1222401265431685
  0.0000000000000000  0.0000000000000000  0.1729013889884428
  0.0000000000000000  0.5000000000000000  0.1729013889884428
  0.2500000000000000  0.2500000000000000  0.1729013889884428
  0.7499999719999977  0.7499999719999977  0.1729013889884428
  0.5000000000000000  0.0000000000000000  0.1729013889884428
  0.7499999719999977  0.2500000000000000  0.1729013889884428
  0.2500000000000000  0.7499999719999977  0.1729013889884428
  0.5000000000000000  0.5000000000000000  0.1729013889884428
  0.5000000000000000  0.2500000000000000  0.2243269895307520
  0.2500000000000000  0.0000000000000000  0.2243269895307520
  0.5000000000000000  0.7499999719999977  0.2243269895307520
  0.2500000000000000  0.5000000000000000  0.2243269895307520
  0.0000000000000000  0.2500000000000000  0.2243269895307520
  0.7499999719999977  0.5000000000000000  0.2243269895307520
  0.0000000000000000  0.7499999719999977  0.2243269895307520
  0.7499999719999977  0.0000000000000000  0.2243269895307520
  0.0000000000000000  0.0000000000000000  0.2756110650176566
  0.7499999719999977  0.7499999719999977  0.2756110650176566
  0.5000000000000000  0.0000000000000000  0.2756110650176566
  0.2500000000000000  0.7499999719999977  0.2756110650176566
  0.2500000000000000  0.2500000000000000  0.2756110650176566
  0.7499999719999977  0.2500000000000000  0.2756110650176566
  0.0000000000000000  0.5000000000000000  0.2756110650176566
  0.5000000000000000  0.5000000000000000  0.2756110650176566
  0.2500000000000000  0.0000000000000000  0.3269200080000019
  0.5000000000000000  0.7499999719999977  0.3269200080000019
  0.0000000000000000  0.7499999719999977  0.3269200080000019
  0.0000000000000000  0.2500000000000000  0.3269200080000019
  0.7499999719999977  0.5000000000000000  0.3269200080000019
  0.2500000000000000  0.5000000000000000  0.3269200080000019
  0.5000000000000000  0.2500000000000000  0.3269200080000019
  0.7499999719999977  0.0000000000000000  0.3269200080000019
  0.0000000000000000  0.0000000000000000  0.3781900079999971
  0.7499999719999977  0.7499999719999977  0.3781900079999971

```
0.0000000000000000 0.5000000000000000 0.3781900079999971
0.5000000000000000 0.5000000000000000 0.3781900079999971
0.5000000000000000 0.0000000000000000 0.3781900079999971
0.2500000000000000 0.2500000000000000 0.3781900079999971
0.2500000000000000 0.7499999719999977 0.3781900079999971
0.7499999719999977 0.2500000000000000 0.3781900079999971
0.0000000000000000 0.2500000000000000 0.4294699980000019
0.7499999719999977 0.0000000000000000 0.4294699980000019
0.5000000000000000 0.7499999719999977 0.4294699980000019
0.2500000000000000 0.5000000000000000 0.4294699980000019
0.7499999719999977 0.5000000000000000 0.4294699980000019
0.2500000000000000 0.0000000000000000 0.4294699980000019
0.0000000000000000 0.7499999719999977 0.4294699980000019
0.5000000000000000 0.2500000000000000 0.4294699980000019
0.5000000000000000 0.5000000000000000 0.4808561835116520
0.2500000000000000 0.7499999719999977 0.4808640512818391
0.0000000000000000 0.5000000000000000 0.4808561835116520
0.7499999719999977 0.2500000000000000 0.4808640512818391
0.0000000000000000 0.0000000000000000 0.4808561835116520
0.7499999719999977 0.7499999719999977 0.4808640512818391
0.2500000000000000 0.2500000000000000 0.4808640512818391
0.5000000000000000 0.0000000000000000 0.4808561835116520
0.2500000000000000 0.0000000000000000 0.5323043819808111
0.5000000000000000 0.2500000000000000 0.5323043819808111
0.2500000000000000 0.5000000000000000 0.5323043819808111
0.0000000000000000 0.7499999719999977 0.5323043819808111
0.7499999719999977 0.0000000000000000 0.5323043819808111
0.0000000000000000 0.2500000000000000 0.5323043819808111
0.5000000000000000 0.7499999719999977 0.5323043819808111
0.7499999719999977 0.5000000000000000 0.5323043819808111
0.2500000000000000 0.7499999719999977 0.5840315775652321
0.5000000000000000 0.5000000000000000 0.5837287475719969
0.5000000000000000 0.0000000000000000 0.5837287475719969
0.7499999719999977 0.2500000000000000 0.5840315775652321
0.2500000000000000 0.2500000000000000 0.5840315775652321
0.0000000000000000 0.5000000000000000 0.5837287475719969
0.7499999719999977 0.7499999719999977 0.5840315775652321
0.0000000000000000 0.0000000000000000 0.5837287475719969
0.2500000000000000 0.5000000000000000 0.6367483763530274
0.5000000000000000 0.2500000000000000 0.6367483763530274
0.7499999719999977 0.0000000000000000 0.6367483763530274
0.0000000000000000 0.7499999719999977 0.6367483763530274
0.7499999719999977 0.5000000000000000 0.6367483763530274
0.0000000000000000 0.2500000000000000 0.6367483763530274
0.5000000000000000 0.7499999719999977 0.6367483763530274
0.2500000000000000 0.0000000000000000 0.6367483763530274
0.0000000000000000 0.0000000000000000 0.1210745471378729
0.5000000000000000 0.0000000000000000 0.1210745471378729
0.5000000000000000 0.5000000000000000 0.1210745471378729
0.2500000000000000 0.2500000000000000 0.1210745471378729
```

```
0.2500000000000000 0.7499999719999977 0.1210745471378729
0.7499999719999977 0.7499999719999977 0.1210745471378729
0.0000000000000000 0.5000000000000000 0.1210745471378729
0.7499999719999977 0.2500000000000000 0.1210745471378729
0.0000000000000000 0.7499999719999977 0.1730970899086515
0.0000000000000000 0.2500000000000000 0.1730970899086515
0.7499999719999977 0.0000000000000000 0.1730970899086515
0.5000000000000000 0.2500000000000000 0.1730970899086515
0.7499999719999977 0.5000000000000000 0.1730970899086515
0.2500000000000000 0.0000000000000000 0.1730970899086515
0.5000000000000000 0.7499999719999977 0.1730970899086515
0.2500000000000000 0.5000000000000000 0.1730970899086515
0.5000000000000000 0.5000000000000000 0.2242964665256295
0.2500000000000000 0.7499999719999977 0.2242964665256295
0.7499999719999977 0.2500000000000000 0.2242964665256295
0.0000000000000000 0.5000000000000000 0.2242964665256295
0.7499999719999977 0.7499999719999977 0.2242964665256295
0.5000000000000000 0.0000000000000000 0.2242964665256295
0.0000000000000000 0.0000000000000000 0.2242964665256295
0.2500000000000000 0.2500000000000000 0.2242964665256295
0.5000000000000000 0.2500000000000000 0.2756174837814882
0.2500000000000000 0.5000000000000000 0.2756174837814882
0.0000000000000000 0.2500000000000000 0.2756174837814882
0.5000000000000000 0.7499999719999977 0.2756174837814882
0.7499999719999977 0.5000000000000000 0.2756174837814882
0.2500000000000000 0.0000000000000000 0.2756174837814882
0.7499999719999977 0.0000000000000000 0.2756174837814882
0.0000000000000000 0.7499999719999977 0.2756174837814882
0.5000000000000000 0.5000000000000000 0.3269200080000019
0.2500000000000000 0.2500000000000000 0.3269200080000019
0.5000000000000000 0.0000000000000000 0.3269200080000019
0.2500000000000000 0.7499999719999977 0.3269200080000019
0.7499999719999977 0.7499999719999977 0.3269200080000019
0.0000000000000000 0.5000000000000000 0.3269200080000019
0.7499999719999977 0.2500000000000000 0.3269200080000019
0.0000000000000000 0.0000000000000000 0.3269200080000019
0.2500000000000000 0.0000000000000000 0.3781900079999971
0.2500000000000000 0.5000000000000000 0.3781900079999971
0.7499999719999977 0.0000000000000000 0.3781900079999971
0.7499999719999977 0.5000000000000000 0.3781900079999971
0.5000000000000000 0.2500000000000000 0.3781900079999971
0.0000000000000000 0.2500000000000000 0.3781900079999971
0.0000000000000000 0.7499999719999977 0.3781900079999971
0.5000000000000000 0.7499999719999977 0.3781900079999971
0.7499999719999977 0.7499999719999977 0.4294699980000019
0.0000000000000000 0.5000000000000000 0.4294699980000019
0.2500000000000000 0.2500000000000000 0.4294699980000019
0.5000000000000000 0.0000000000000000 0.4294699980000019
0.0000000000000000 0.0000000000000000 0.4294699980000019
0.7499999719999977 0.2500000000000000 0.4294699980000019
```

```
0.5000000000000000 0.5000000000000000 0.4294699980000019
0.2500000000000000 0.7499999719999977 0.4294699980000019
0.5000000000000000 0.2500000000000000 0.4807581974299096
0.2500000000000000 0.0000000000000000 0.4807581974299096
0.7499999719999977 0.5000000000000000 0.4807581974299096
0.7499999719999977 0.0000000000000000 0.4807581974299096
0.0000000000000000 0.7499999719999977 0.4807581974299096
0.5000000000000000 0.7499999719999977 0.4807581974299096
0.2500000000000000 0.5000000000000000 0.4807581974299096
0.0000000000000000 0.2500000000000000 0.4807581974299096
0.0000000000000000 0.0000000000000000 0.5320284580287068
0.7499999719999977 0.2500000000000000 0.5320030882989003
0.0000000000000000 0.5000000000000000 0.5320284580287068
0.2500000000000000 0.7499999719999977 0.5320030882989003
0.7499999719999977 0.7499999719999977 0.5320030882989003
0.5000000000000000 0.0000000000000000 0.5320284580287068
0.2500000000000000 0.2500000000000000 0.5320030882989003
0.5000000000000000 0.5000000000000000 0.5320284580287068
0.7499999719999977 0.5000000000000000 0.5832505892617470
0.0000000000000000 0.2500000000000000 0.5832505892617470
0.0000000000000000 0.7499999719999977 0.5832505892617470
0.7499999719999977 0.0000000000000000 0.5832505892617470
0.5000000000000000 0.2500000000000000 0.5832505892617470
0.2500000000000000 0.0000000000000000 0.5832505892617470
0.5000000000000000 0.7499999719999977 0.5832505892617470
0.2500000000000000 0.5000000000000000 0.5832505892617470
0.7499999719999977 0.7499999719999977 0.6333524811864069
0.0000000000000000 0.5000000000000000 0.6347510236623711
0.7499999719999977 0.2500000000000000 0.6333524811864069
0.0000000000000000 0.0000000000000000 0.6347510236623711
0.5000000000000000 0.0000000000000000 0.6347510236623711
0.2500000000000000 0.7499999719999977 0.6333524811864069
0.5000000000000000 0.5000000000000000 0.6347510236623711
0.2500000000000000 0.2500000000000000 0.6333524811864069
0.2500000000000000 0.0000000000000000 0.6879949475930847
0.7500000000000000 0.0000000000000000 0.6879949475930847
0.2500000000000000 0.5000000000000000 0.6879949475930847
0.7500000000000000 0.5000000000000000 0.6879949475930847
0.0000000000000000 0.2500000000000000 0.6879949475930847
0.0000000000000000 0.7500000000000000 0.6879949475930847
0.5000000000000000 0.2500000000000000 0.6879949475930847
0.5000000000000000 0.7500000000000000 0.6879949475930847
0.0000000000000000 0.0000000000000000 0.6934589732240349
0.5000000000000000 0.5000000000000000 0.6934589732240349
0.5000000000000000 0.0000000000000000 0.6934589732240349
0.0000000000000000 0.5000000000000000 0.6934589732240349
```